\documentclass[10pt]{iopart}

\usepackage{iopams}
\usepackage{iopams}
\usepackage{hyperref}
\hypersetup{colorlinks=true, urlcolor=red, citecolor=blue, linkcolor=blue}
\usepackage{epsfig}
\usepackage{newlfont}
\usepackage{amssymb}
\usepackage{bm}
\usepackage{subfigure}

\def\lsim{\mathrel{\rlap{\lower4pt\hbox{\hskip1pt$\sim$}}
    \raise1pt\hbox{$<$}}}                
\def\gsim{\mathrel{\rlap{\lower4pt\hbox{\hskip1pt$\sim$}}
    \raise1pt\hbox{$>$}}}                

\begin{document}

\renewcommand*{\DefineNamedColor}[4]{%
   \textcolor[named]{#2}{\rule{7mm}{7mm}}\quad
  \texttt{#2}\strut\\}

\definecolor{red}{rgb}{1,0,0}
\definecolor{cyan}{cmyk}{1,0,0,0}

\title[Generating continuous variable entangled states...]{Generating continuous variable entangled states for quantum teleportation using a superposition of number-conserving operations}

\author{Himadri Shekhar Dhar\(^{1,2}\), Arpita Chatterjee\(^2\), and Rupamanjari Ghosh\(^{2,3}\)}

\address{\(^1\)Harish-Chandra Research Institute, Chaatnag Road, Jhunsi, Allahabad 200019, India\\
\(^2\)School of Physical Sciences, Jawaharlal Nehru University, New Delhi 110067, India\\
\(^3\)School of Natural Sciences, Shiv Nadar University, Gautam Buddha Nagar, UP 201314, India.
}
\ead{rghosh.jnu@gmail.com}

\date{\today}

\begin{abstract}
We investigate the states generated in continuous variable (CV) optical fields on operating them with a number-conserving operator of the type $s\hat{a}\hat{a}^\dag + t\hat{a}^\dag\hat{a}$, formed by the generalised superposition of products of field annihilation ($\hat{a}$) and creation ($\hat{a}^\dag$) operators, with $s^2+t^2=1$. Such an operator is experimentally realizable and can be suitably manipulated to generate nonclassical optical states when applied on single- and two-mode coherent, thermal, and squeezed input states. At low intensities, these nonclassical states can interact with a secondary mode via a linear optical device to generate two-mode discrete entangled states, which can serve as a resource in quantum information protocols. The advantage of these operations are tested by applying the generated entangled states as quantum channels in CV quantum teleportation, under the Braunstein and Kimble protocol. We observe that, under these operations, while the average fidelity of CV teleportation is enhanced for the nonclassical channel formed using input squeezed states, it remains at the classical threshold for input coherent and thermal states. This is due to the fact that though these operations can introduce discrete entanglement in all input states, it enhances the Einstein-Podolosky-Rosen (EPR) correlations only in the nonclassical squeezed state inputs, leading to an advantage in CV teleportation. This shows that nonclassical optical states generated using the above operations on classical coherent and thermal state inputs are not resourceful for CV teleportation. This investigation could prove useful in efficient implementation of noisy non-Gaussian channels, formed by linear operations, in future teleportation protocols. 
\end{abstract}

(Some figures may appear in colour only in the online journal)


\submitto{\JPB}
\maketitle

\section{Introduction}
\label{ss1}

The wide use of Gaussian and non-Gaussian electromagnetic field states for various quantum tasks and protocols renders the generation and investigation of these quantum states an essential aspect of modern research. Entangled states generated using both linear and nonlinear quantum optical operations have been extensively used in quantum information \cite{Braun} and communications \cite{Brieg} applications. The accessibility of continuous variable (CV) processing in quantum information theory has led to successful implementation of novel quantum tasks such as quantum teleportation \cite{vaid, BK}, quantum cryptography \cite{crypto} and quantum memory \cite{mem, mem1} (for reviews, see \cite{Cerf, Japan, Lloyd}).

A recent method of generating nonclassicality in electromagnetic field states is by the operation of photon addition ($\hat{a}^\dag$) \cite{add} and subtraction ($\hat{a}$) \cite{subtract}. A remarkable aspect of the non-Gaussian photon addition and subtraction operations is the relative efficiency with which these operations can be experimentally realised using linear optical devices and parametric down-converters \cite{expt}. These non-Gaussian operations are known to enhance entanglement \cite{enhanceent} and quantum information protocols, such as quantum teleportation \cite{enhanceqi}. An extension of these non-Gaussian operations, such as photon addition followed by photon subtraction or vice-versa, has also been implemented to show improvements in quantum correlations and other quantum tasks \cite{refyang, refcerf}. The study of nonclassical optical states generated by such non-Gaussian operations has received considerable theoretical attention in quantum optics \cite{opth, arpita}. Further, the experimental implementation of these operations has been used to prove the canonical commutation relation \cite{kim}. A useful operation based on the non-Gaussian photon-addition and subtraction protocols is the number-conserving generalised superposition of products (GSP) of field annihilation ($\hat{a}$) and creation ($\hat{a}^\dag$) operators of the form $s\hat{a}\hat{a}^\dag + t\hat{a}^\dag\hat{a}$, with $s^2+t^2=1$. Such an operation is a generalization of the experimental scheme proposed to study the bosonic canonical commutation in optical fields \cite{kim}. Using an analysis of quasi-probability functions, quadrature squeezing and sub-Poissonian statistics, we have recently shown that such GSP operations introduces nonclassicality in single-mode coherent and thermal states \cite{arpita}. 

The experimental scheme to implement the GSP operation using linear optical devices and parametric down-converters has been illustrated and mathematically explained in Ref.~\cite{arpita}. However, the feasibility of such a scheme is reliant on the efficiency with which photon addition and subtraction can be experimentally achieved. In particular, it is limited by the low efficiency with which a single photon can be detected in the relevant modes \cite{KIMnew} (cf. \cite{add,subtract,expt}).

It is known that GSP operations can be suitably manipulated to generate nonclassical output states when applied on single- and two-mode coherent, thermal and squeezed vacuum states \cite{arpita}. Hence, the primary motivation of the present article is to analyze the nonclassicality introduced by the GSP operations from a quantum information perspective and investigate if the generated nonclassicality can enhance the performance of specific quantum information tasks.
These nonclassical optical states, in mode \emph{a}, can then be converted into two-mode entangled states by interacting with a secondary mode \emph{b}, via a linear optical device such as a 50:50 beam-splitter. For a single-mode input state, the secondary mode is the vacuum, while for a two-mode input state, one mode serves as \emph{a} and the other mode as \emph{b} (see figure \ref{fig1}). We note that the generated  entanglement is characteristic of the nonclassicality of the GSP operated state, as beam-splitters do not generate entanglement for classical inputs \cite{kimbs}. Thus the generated entanglement is an indicator of the resource obtained due to the GSP operations. 
The suitability of the two-mode entangled state is investigated for potential use as a quantum channel in a CV quantum teleportation, under the Braunstein and Kimble (BK) protocol \cite{BK}. The success of the protocol is evaluated in terms of the average fidelity obtained in teleporting a single-mode optical state. The efficiency of the quantum channel for CV teleportation can be justified in terms of the phase-quadrature or the Einstein-Podolsky-Rosen (EPR) correlations present in the quantum channels.

The main results obtained in our investigation can be summarized in the following points:~1) At low field intensities, the GSP operation can be suitably tuned to generate highly entangled two-mode discrete level optical states for both single- and two-mode coherent, thermal and squeezed input states. This is a signature of the nonclassicality generated by the GSP operation on these input states.~2) These two-mode entangled states can be used as quantum channels in CV quantum teleportation of single-mode coherent and squeezed states, under the BK protocol.~3) We observe that while the average fidelity of CV teleportation is enhanced for the squeezed input states under GSP operations, the fidelity does not rise above the classical threshold for coherent and thermal inputs. This shows that the generated nonclassicality in coherent and thermal states, under GSP operations, is not resourceful for CV teleportation.~4) Investigating further, we observe that the GSP operations do not enhance the nonclassicality of coherent and thermal input states in terms its phase-quadrature or the EPR correlations, which is crucial for gaining an advantage in CV teleportation. Interestingly, the above results show that though GSP operations introduce nonclassicality and discrete level entanglement for all input states, the injected nonclassicality is not resourceful for CV teleportation when the input state is coherent or thermal, as it fails to enhance the EPR correlations. The real advantage is observed only when the input state is already nonclassical, such as the squeezed state. This result is important from the consideration of noisy quantum channels, where classical noise can be modelled using coherent or thermal states.

The paper is organised as follows. The introduction in section \ref{ss1} is followed by a discussion on the characterization of nonclassicality and two-mode entanglement in section \ref{ent}. We then discuss the state generation protocol for single-mode (section \ref{ss2}) and two-mode (section \ref{ss3}) coherent, thermal and squeezed state inputs. We investigate the entanglement properties in section \ref{ss5} followed by the investigation of average teleportation fidelities in section \ref{ss6}. In section \ref{sec-epr}, we look at the EPR correlations of the quantum channels, and end with a discussion of the results in section \ref{discuss}.

\section{Measure of nonclassicality}
\label{ent}

The action of number-conserving operators formed by the GSP of field annihilation ($\hat{a}$) and creation ($\hat{a}^\dag$) operators of the type $s\hat{a}\hat{a}^\dag + t\hat{a}^\dag\hat{a}$, with $s^2+t^2=1$, on field modes introduces distinct nonclassicality in the operated states \cite{arpita}. The nonclassicality in these cases is characterized by the indicators such as negative quasiprobability distributions, sub-Poissonian statistics or quadrature squeezing. Under such a description, coherent and thermal states can be deemed as classical, wheras squeezed vacuum states are highly nonclassical.
From the perspective of quantum information theory, the nonclassicality of the GSP operated state can be investigated in terms of the two-mode entanglement that develops upon interaction of the modes via some linear optical device. Nonclassical states upon interacting with a secondary mode via a beam splitter will develop finite entanglement \cite{kimbs}. For example, non-Gaussian operations based on photon addition and subtraction are known to enhance entanglement in squeezed vacuum states \cite{refyang}.

\begin{figure}
\vspace{0.2cm}
\begin{center}
\epsfig{figure=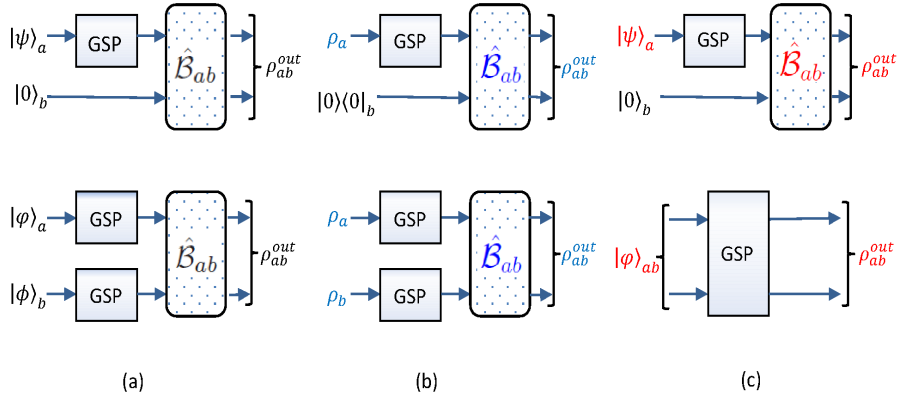
,width=.9\textwidth,height=0.25\textheight}
\end{center}
\caption{Schematic of the generation of bipartite entangled output with single-mode state preparation (top row) and two-mode state preparation (bottom row) for (a) coherent input state, (b) thermal input state, and (c) squeezed input state. $\hat{\mathcal{B}}_{ab}$ represents the beam-splitter operation. The experimental scheme for performing a generalised superposition of products (GSP) operation is given in Ref.\,\cite{arpita}.}
\label{fig1}
\end{figure}

The protocol adopted to measure the nonclassicality using entanglement is different for the case of single-mode and two-mode inputs. For a single-mode GSP operated state, in mode \emph{a}, we interact the single-mode output with a vacuum in mode \emph{b}, via a 50:50 beam splitter to generate a two-mode output state. A schematic for the GSP protocol and the beam-splitter interaction for single-mode input states is given in the upper row of figure \ref{fig1}. The entanglement in this two-mode output state is measured using logarithmic negativity (LN) \cite{LN}, which gives us a measure of the entanglement capacity or potential \cite{EP} of the GSP operated state. We call this measure entanglement capacity (EC), as it measures the capacity of the nonclassical GSP operated state to create entanglement via a linear optical device, and is thus a measure of the nonclassicality of the GSP operated state. The EC is zero in cases where a classical input state interacts with the vacuum mode via a beam-splitter \cite{kimbs}.
%
%
%
%
%

For two-mode input states which are initially ``classical''\footnote{These quantum states are ``classical'' in the sense that they are closest to a classical optical field, in terms of a positive quasiprobability distributions, Poissonian photon statistics and zero squeezing in quadrature variables.}, such as the coherent and thermal states, the GSP operation is applied separately on each mode. This operation does not introduce any entanglement in the two-mode output since the GSP operations are local. However, the two GSP operated modes are individually nonclassical and this can again be converted into two-mode entanglement using a linear optical device, which introduces a nonlocal interaction.
To calculate the EC for the two-mode GSP operated state, we allow the two modes to interact via a 50:50 beam splitter to generate entangled two-mode output states. The EC is then calculated using LN. For two-mode input states which are initially entangled, such as a two-mode squeezed state, the GSP operation is globally applied to the two modes without further interaction between the modes.
A schematic for the complete protocol for two-mode input states is given in the lower row of figure \ref{fig1}.

We briefly describe the entanglement measure called logarithmic negativity that is used to characterize the entanglement capacity. LN is defined \cite{LN} on the known premise that the negativity of the partial transpose is a sufficient condition for a bipartite quantum state to be entangled. This is the Peres-Horodecki separability criterion \cite{10a}. For low-dimensional bipartite systems (qubit-qubit and qubit-qutrit) the negativity of partial transpose is not only sufficient but also a necessary condition for bipartite entanglement \cite{sepcriteria}.
To calculate LN, we evaluate a quantity called ``negativity'', $\cal{N}(\rho_{ab}) = (||\rho_{ab}^{T_a}||_1- 1)/2$, where $||\rho_{ab}^{T_a}||_1$ is the trace norm of the partial transpose \(\rho_{ab}^{T_a}\), which is positive for all non-entangled states. Hence $\cal{N}(\rho_{AB})$ is zero for separable states.
LN of \(\rho_{AB}\) is defined as
\begin{equation}
E_{\cal{N}}(\rho_{AB}) =  \log_2 \left\|\rho_{AB}^{T_A}\right\|_1 \equiv \log_2 \left[ 2 \cal{N}(\rho_{AB}) +1 \right] .
\label{logneg}
\end{equation}
We use LN as the measure of two-mode entanglement to calculate EC for all GSP operated states.

\section{GSP operated output states}
\label{ss}

In this section, we consider the number-conserving generalized superposition operation of the form $s\hat{a}\hat{a}^\dag + t\hat{a}^\dag\hat{a}$, with $s^2+t^2=1$, acting on single- and two-mode coherent, thermal and squeezed vacuum states. We mathematically derive the relations to convert the generated nonclassical state into entangled two-mode states, by using a linear optical device, such as a 50:50 beam-splitter, as per the protocol formalised in section \ref{ent}. The final output state obtained is then varied with the system parameter $s$, to analyze the entanglement capacity of these states and their utility in CV quantum teleportation in the following sections.

\subsection{Single-mode input states}
\label{ss2}
\vspace{0.25cm}
\noindent\emph{Single-Mode Coherent State}: Let us consider the GSP operation acting on a single-mode coherent state $|\psi\rangle_a \equiv |\alpha\rangle_a$ [figure \ref{fig1}(a), top row]. The single-mode (in mode $a$) GSP operated coherent state is given by
\begin{eqnarray}
|\alpha'\rangle_a &=& \frac{1}{\sqrt{N}}(s\hat{a}\hat{a}^\dag + t\hat{a}^\dag\hat{a})|\alpha\rangle_a \nonumber \\
&=& \frac{1}{\sqrt{N}}(s|\alpha\rangle_a + (s+t) ~\alpha \hat{a}^\dag |\alpha\rangle_a) ,
\label{1}
\end{eqnarray}
where $N$ is the normalization constant. In terms of the displacement operator, $D_a(\alpha) = \exp(\hat{a}^\dag\alpha-\hat{a}\alpha^*)$, a coherent state can be written as $|\alpha\rangle_a = D_a(\alpha)|0\rangle_a$ \cite{Scully}. Hence we can write relation (\ref{1}) as
\begin{eqnarray}
|\alpha'\rangle_a 
&=& \frac{D_a(\alpha)}{\sqrt{N}}(s + \alpha (s+t) (\hat{a}^\dag + \alpha^*))|0\rangle_a \nonumber \\
&=& \frac{D_a(\alpha)}{\sqrt{N}}(s|0\rangle_a + \alpha (s+t)[~|1\rangle_a+\alpha^*|0\rangle_a~]),
\label{2}
\end{eqnarray}
where we have used the relation $D_a^\dag(\alpha)\hat{a}^\dag D_a(\alpha)= \hat{a}^\dag+\alpha^*$.
The single-mode GSP operated coherent state is then interacted with a vacuum mode via a 50:50 beam splitter.
The two-mode output state after the interaction has the form
$|\phi\rangle_{ab}^{out}= \hat{\mathcal{B}}_{ab} ~ |\alpha'\rangle_a \otimes |0\rangle_b$,
where $\hat{\mathcal{B}}_{ab} = \exp(\pi/2\{\hat{a}^\dag \hat{b}-\hat{a}\hat{b^\dag}\})$ is the beam-splitter operator with input ports $a$ and $b$ [figure \ref{fig1}(a), top row]. Using relation (\ref{2}), we obtain
\begin{eqnarray}
|\phi\rangle_{ab}^{out} 
&=& \frac{\hat{\mathcal{B}}_{ab}D_a(\alpha)\hat{\mathcal{B}}_{ab}^\dag}{\sqrt{N}}\hat{\mathcal{B}}_{ab}\left[p_1 |0\rangle_a|0\rangle_b  + p_2 \hat{a}^\dag |0\rangle_a|0\rangle_b \right] ,
\label{5}
\end{eqnarray}
where $p_1=s+(s+t)|\alpha|^2$ and $p_2=\alpha(s+t)$. Using the fact that $\hat{\mathcal{B}}_{ab} D_a(\alpha) \hat{\mathcal{B}}_{ab}^\dag$ = $ D_a(\alpha/\sqrt{2})D_b(-\alpha/\sqrt{2})$, and the beam-splitter relations, $\hat{\mathcal{B}}_{ab} \hat{a}^\dag \hat{\mathcal{B}}_{ab}^\dag$ = $\frac{1}{\sqrt{2}}(\hat{a}^\dag-\hat{b}^\dag)$ and $\hat{\mathcal{B}}_{ab} \hat{b}^\dag \hat{\mathcal{B}}_{ab}^\dag$ = $\frac{1}{\sqrt{2}}(\hat{b}^\dag+\hat{a}^\dag)$, we obtain
\begin{eqnarray}
|\phi\rangle_{ab}^{out} 
&=& \mathcal{A}_{ab} \left[ p_1  + p_2/\sqrt{2} (\hat{a}^\dag - \hat{b}^\dag)  \right]|0\rangle_a|0\rangle_b
\nonumber \\
&=& \mathcal{A}_{ab} \left[ p_1 |00\rangle_{ab} + \frac{p_2}{\sqrt{2}}(|10\rangle_{ab} - |01\rangle_{ab}) \right] ,
\label{6}
\end{eqnarray}
where $\mathcal{A}_{ab}$=$ D_a(\alpha/\sqrt{2})D_b(-\alpha/\sqrt{2})/\sqrt{N}$, $|ij\rangle_{ab}$=$|i\rangle_a|j\rangle_b$.

The density matrix for the two-mode state given by relation (\ref{6}) is $\rho_{ab}^{out} = \mathcal{A}_{ab} ~\rho_{ab}^0~\mathcal{A}_{ab}^\dag$, where $\rho_{ab}^0$ is given by
\begin{equation}
\rho_{ab}^0 = \left( \begin{array}{cccc}
p_1^2 & -p_1p_2/\sqrt{2} & p_1p_2/\sqrt{2} & 0 \\
-p_1p_2/\sqrt{2} & p_2^2/2 & -p_2^2/2 & 0 \\
p_1p_2/\sqrt{2} & -p_2^2/2 &  p_2^2/2 & 0 \\
0 & 0 & 0 & 0 \end{array} \right) .
\label{7}
\end{equation}
The operator $\mathcal{A}_{ab}$ acts locally on the modes $a$ and $b$, and hence cannot change the entanglement properties of the two-qubit density matrix (\ref{7}). This reduces an infinite-dimensional two-mode interaction (in an infinite dimensional Fock state basis) to a discrete two-qubit problem (in the \{$|0\rangle, |1\rangle$\} basis). Hence, the entanglement capacity can simply be calculated by computing the LN of the smaller two-qubit state. Hence, the GSP operation generates a nonclassical two-mode output state that can be used in possible quantum protocols. \\



\noindent\emph{Single-Mode Thermal State}: A similar approach can be taken if the single-mode input is a classical Gaussian thermal state. The thermal state is a maximally mixed state in the Fock state basis, and the thermal density matrix can be represented as
\begin{equation}
\rho^{th}_a=\frac{1}{1+\bar{n}}\sum_{n=0}^{\infty}\left(\frac{\bar{n}}{1+\bar{n}}\right)^n |n\rangle\langle n|_a ,
\label{8}
\end{equation}
where $\bar{n}$ is the average photon number. The action of GSP on the thermal field density matrix gives us the following operated state \cite{arpita}:
\begin{equation}
\rho^{'~th}_a = \frac{M^{-1}}{1+\bar{n}}\sum_{n=0}^{\infty}\left(\frac{\bar{n}}{1+\bar{n}}\right)^n (s+n(s+t))^2 |n\rangle\langle n|_a ,
\label{9}
\end{equation}
where $M$ is the normalization constant. The GSP operated thermal state (\ref{9}) is thus a mixed state in the infinite dimensional basis. It is known that the GSP operated single-mode thermal state has nonclassicality introduced by the operation and is non-Gaussian \cite{arpita}.
The GSP operated state (in mode $a$) is interacted with a vacuum state in mode $b$ via a 50:50 beam splitter [figure \ref{fig1}(b), top row] to generate a nonclassical two-mode output state. The operated two-mode state is of the form
\begin{eqnarray}
\rho_{ab} &=& \hat{\mathcal{B}}_{ab}~ \left(\rho^{'~th}_a \otimes |0\rangle\langle 0|_b \right)~ \hat{\mathcal{B}}_{ab}^\dag \nonumber\\
&=& \hat{\mathcal{B}}_{ab} \left(\sum_{n=0}^{\infty} \frac{q_n}{n!} \left(\hat{a}^\dag\right)^n|0 0\rangle\langle 0 0|_{ab} \left(\hat{a}\right)^n \right) \hat{\mathcal{B}}_{ab}^\dag ,
\label{10}
\end{eqnarray}
where $q_n=\left(\frac{M^{-1}}{1+\bar{n}}\right)\left(\frac{\bar{n}}{1+\bar{n}}\right)^n \left( s + n \left(s+t\right)\right)^2$. The infinite-dimensional interaction between the two modes can be reduced by considering a truncated thermal state input with a low average photon number. For $\bar{n} \leq$ 0.1, the two-mode density matrix (\ref{10}) can be written as
\begin{eqnarray}
\rho_{ab} &=& \hat{\mathcal{B}}_{ab} \left(q_0|0 0\rangle\langle 0 0|_{ab} + q_1 \hat{a}^\dag|0 0\rangle\langle 0 0|_{ab} \hat{a} \right. \nonumber\\
&&+ \left. \frac{q_2}{2} (\hat{a}^\dag)^2|0 0\rangle\langle 0 0|_{ab} (\hat{a})^2\right) \hat{\mathcal{B}}_{ab}^\dag ,
\label{11}
\end{eqnarray}
where only the first three terms ($n =$ 0, 1, 2) in the summation have been retained after truncation.
Applying the unitary operation $\hat{\mathcal{B}}_{ab}$, we get the final two-mode output state as
\begin{eqnarray}
\rho_{ab}^{out} &=& q_0 |00\rangle\langle 00| + q_1 \hat{\mathcal{B}}_{ab} \hat{a}^\dag \hat{\mathcal{B}}_{ab}^\dag |00\rangle \langle 00| \hat{\mathcal{B}}_{ab} \hat{a} \hat{\mathcal{B}}_{ab}^\dag \nonumber \\
&& + q_2 \hat{\mathcal{B}}_{ab} (\hat{a}^\dag)^2 \hat{\mathcal{B}}_{ab}^\dag |00\rangle \langle 00| \hat{\mathcal{B}}_{ab} (\hat{a})^2 \hat{\mathcal{B}}_{ab}^\dag \nonumber\\
&=& q_0 |00\rangle\langle 00| + \frac{q_1}{2}\left(\hat{a}^\dag-\hat{b}^\dag\right)|00\rangle \langle 00| \left(\hat{a}-\hat{b}\right)\nonumber \\
&& + \frac{q_2}{4}\left(\hat{a}^\dag-\hat{b}^\dag\right)^2|00\rangle \langle 00| \left(\hat{a}-\hat{b}\right)^2 .
\label{12}
\end{eqnarray}
The final density matrix can be written as
\begin{equation}
\rho_{ab}^{out} = \left( \begin{array}{ccccccccc}
q_0 & 0 & 0 & 0 & 0 & 0 & 0 & 0 & 0\\
0 & \frac{q_1}{2} & 0 & \frac{q_1}{2} & 0 & 0 & 0 & 0 & 0 \\
0 & 0 & \frac{q_2}{4} & 0 &  -\frac{q_2}{2\sqrt{2}} & 0 & \frac{q_2}{4} & 0 & 0 \\
0 & \frac{q_1}{2} & 0 & \frac{q_1}{2} & 0 & 0 & 0 & 0 & 0 \\
0 & 0 & -\frac{q_2}{2\sqrt{2}} & 0 &  \frac{q_2}{2} & 0 & -\frac{q_2}{2\sqrt{2}} & 0 & 0 \\
0 & 0 & 0 & 0 & 0 & 0 & 0 & 0 & 0 \\
0 & 0 & \frac{q_2}{4} & 0 &  -\frac{q_2}{2\sqrt{2}} & 0 & \frac{q_2}{4} & 0 & 0 \\
0 & 0 & 0 & 0 & 0 & 0 & 0 & 0 & 0 \\
0 & 0 & 0 & 0 & 0 & 0 & 0 & 0 & 0 \end{array} \right) .
\label{13}
\end{equation}
%
The low average photon number ensures that the density matrix is effectively truncated to a discrete $\mathit{3^{\otimes 2}}$-dimensional two-mode system. Hence, the GSP operation generates a discrete two-mode nonclassical state.\\


\noindent\emph{Single-Mode Squeezed State}: The GSP operation and subsequent conversion to a two-mode entangled state for a single-mode squeezed state are similar to the approach adopted for coherent and thermal input states. However, the squeezed state is not a classical input state unlike the cases of coherent and thermal inputs. The single-mode (in mode $a$) squeezed state can be written as
\begin{eqnarray}
\mathcal{S}(z)|0\rangle &=& \exp[(z/2)(\hat{a}^2-\hat{a}^{\dag 2})]|0\rangle_a \nonumber \\
&=& (1-\lambda^2)^{1/4}\sum_{n=0}^\infty \frac{\sqrt{(2n)!}}{n!}\left(-\frac{\lambda}{2}\right)^n|2n\rangle_a ,
\label{sq1.1}
\end{eqnarray}
where $\mathcal{S}(z)$ is the squeezing operator, $z$ is the squeezing parameter, and $\lambda \equiv \tanh z$. The GSP operated squeezed state [figure \ref{fig1}(c), top row] is given by
\begin{eqnarray}
|\psi\rangle &=& \frac{1}{\sqrt{N}}(s\hat{a}\hat{a}^\dag + t\hat{a}^\dag\hat{a})\mathcal (1-\lambda^2)^{1/4}
\sum_{n=0}^\infty \frac{\sqrt{(2n)!}}{n!}\left(-\frac{\lambda}{2}\right)^n|2n\rangle_a \nonumber\\
&=& \sqrt{\frac{(1-\lambda^2)^{1/2}}{N}}\sum_{n=0}^\infty \frac{\sqrt{(2n)!}}{n!}\left(-\frac{\lambda}{2}\right)^n (s+2n(s+t))|2n \rangle_a ,
\label{sq1.2}
\end{eqnarray}
where $N$ is the normalization constant. The GSP operated single-mode squeezed state (in mode $a$) is interacted with a vacuum state (in mode $b$) via a 50:50 beam splitter. The resulting two-mode state is given by
\begin{equation}
|\psi'\rangle = \hat{\mathcal{B}}_{ab}\sum_{n=0}^\infty c_n \frac{(\hat{a}^\dag)^{2n}}{\sqrt{(2n)!}}|0\rangle_a|0\rangle_b ,
\label{sq1.3}
\end{equation}
where $|0\rangle$ is the vacuum mode, $\hat{\mathcal{B}}_{ab}$ is the beam-splitter operator, and $c_n \equiv \sqrt{\frac{(1-\lambda^2)^{\frac{1}{2}}}{N}}\frac{\sqrt{(2n)!}}{n!}\left(-\frac{\lambda}{2}\right)^n (s+2n(s+t))$.

The infinite-dimensional two-mode output can be truncated by considering regimes where $\lambda \ll$ 1. In the low $\lambda$ regime, the GSP operated single-mode squeezed state can be written as
\begin{equation}
|\psi'\rangle = \hat{\mathcal{B}}_{ab}(c_0 |0\rangle_a|0\rangle_b + c_1 (\hat{a}^\dag)^2|0\rangle_a|0\rangle_b) ,
\label{sq1.4}
\end{equation}
where $c_0 = \sqrt{\frac{(1-\lambda^2)^{\frac{1}{2}}}{N}} s$, and $c_1 = -\sqrt{\frac{(1-\lambda^2)^{\frac{1}{2}}}{N}}\lambda (\frac{s}{2}+(s+t))$. Applying the beam-splitter operation, we get
\begin{eqnarray}
|\psi'\rangle &=& c_0 |00\rangle_{ab} + c_1 \hat{\mathcal{B}}_{ab} (\hat{a}^\dag)^2\hat{\mathcal{B}}_{ab}^\dag|00\rangle_{ab} \nonumber\\
&=& c_0 |00\rangle_{ab} + c_1\left(\frac{\hat{a}^\dag-\hat{b}^\dag}{\sqrt{2}}\right)\left(\frac{\hat{a}^\dag-\hat{b}^\dag}{\sqrt{2}}\right)|00\rangle_{ab}\nonumber\\
&=& c_0 |00\rangle_{ab} + \frac{c_1}{\sqrt{2}} |20\rangle_{ab}+\frac{c_1}{\sqrt{2}}|02\rangle_{ab}-c_1|11\rangle_{ab}. \nonumber\\
\label{sq1.5}
\end{eqnarray}

The output two-mode density matrix is given by $\rho_{ab}^{out} = |\psi'\rangle\langle \psi'|$. The GSP operated two-mode output can be used as a quantum channel in CV teleportation.

\subsection{Two-mode input states}
\label{ss3}

\vspace{0.25cm}
\noindent\emph{Two-Mode Coherent State}: For a classical coherent state input, the GSP operation is applied separately on the two modes $a$ and $b$ [figure \ref{fig1}(a), bottom row]. Each mode operation is similar to the case for a single-mode input. The two modes than interact via the beam-splitter. The single-mode GSP operated input coherent state is given by relation (\ref{2}). The beam-splitter interaction for the two modes can then be written as $|\phi \rangle_{ab}^{out} = \hat{\mathcal{B}}_{ab} \left( |\alpha' \rangle_a \otimes |\beta' \rangle_b \right)$, where $|\alpha\rangle_a$ and $|\beta\rangle_b$ are the two input coherent states. The output state can then be written as
\begin{eqnarray}
|\phi \rangle_{ab}^{out} 
&=& \hat{\mathcal{B}}_{ab} \frac{D_a(\alpha)D_b(\beta)}{\sqrt{N_1 N_2}} \left[\left( s_1|0\rangle_a \right. \right. 
\left. + (s_1+t_1) \alpha \left( |1\rangle_a + \alpha^* |0\rangle_a \right)\right) \left(s_2|0\rangle_b \right. \nonumber\\ 
&& + \left. \left. (s_2+t_2) \beta ~ \left( |1\rangle_b + \beta^* |0\rangle_b \right)\right)\right] ,\nonumber\\
&=& \mathcal{A}_{ab}^{local}\left[ p'_1 |0\rangle_{a}|0\rangle_{b} + p'_2 ~\hat{\mathcal{B}}_{ab}\hat{b}^\dag \hat{\mathcal{B}}_{ab}^\dag |0 \rangle_a |0 \rangle_b \right. + p'_3 \hat{\mathcal{B}}_{ab}\hat{a}^\dag \hat{\mathcal{B}}_{ab}^\dag |0 \rangle_a |0 \rangle_b \nonumber\\ && \left. + p'_4 ~\hat{\mathcal{B}}_{ab}\hat{a}^\dag\hat{b}^\dag \hat{\mathcal{B}}_{ab}^\dag |0 \rangle_a |0 \rangle_b \right] ,
\label{14}
\end{eqnarray}
where
\begin{eqnarray}
\mathcal{A}_{ab}^{local} &=& \frac{1}{\sqrt{N_1 N_2}} \times  D_a\left(\frac{\alpha}{\sqrt{2}}\right)D_b\left(\frac{-\alpha}{\sqrt{2}}\right)D_a\left(\frac{\beta}{\sqrt{2}}\right)D_b\left(\frac{\beta}{\sqrt{2}}\right) , \nonumber
\end{eqnarray}
and
\begin{eqnarray}
p'_1 &=& s_1 s_2 + s_1(s_2+t_2)|\beta|^2 + s_2(s_1+t_1)|\alpha|^2 \nonumber\\
&& + (s_1+t_1)(s_2+t_2)|\alpha|^2|\beta|^2 , \nonumber \\
p'_2 &=& s_1(s_2+t_2)\beta + (s_1+t_1)(s_2+t_2)|\alpha|^2 \beta , \nonumber\\
p'_3 &=& s_2(s_1+t_1)\alpha + (s_1+t_1)(s_2+t_2)|\beta|^2 \alpha , \nonumber\\
p'_4 &=& \alpha \beta (s_1+t_1)(s_2+t_2) . \nonumber
\end{eqnarray}
Hence, applying the beam-splitter operation on equation (\ref{14}), we obtain
\begin{eqnarray}
|\phi \rangle_{ab}^{out} &=& \mathcal{A}_{ab}^{local}\left[p'_1 + p'_2 \frac{\hat{b}^\dag+\hat{a}^\dag}{\sqrt{2}} + p'_3 \frac{\hat{a}^\dag -\hat{b}^\dag}{\sqrt{2}} \right. \nonumber\\
&& + \left. p'_4 \left(\frac{\hat{a}^\dag-\hat{b}^\dag}{\sqrt{2}}\right)\left( \frac{\hat{b}^\dag+\hat{a}^\dag}{\sqrt{2}}\right)\right]|00\rangle_{ab} \nonumber\\ \nonumber\\
&=& \mathcal{A}_{ab}^{local}\left[p''_1|00\rangle_{ab} + p''_2 |01\rangle_{ab} + p''_3|10\rangle_{ab} \right. \nonumber\\ \nonumber\\
&& - \left. p''_4 |02\rangle_{ab} +p''_4|20\rangle_{ab}\right] ,
\label{16}
\end{eqnarray}
where $p''_1=p'_1$, $p''_2=\frac{p'_2-p'_3}{\sqrt{2}}$, $p''_3=\frac{p'_2+p'_3}{\sqrt{2}}$ and $p''_4=\frac{p'_4}{\sqrt{2}}$.

The two-mode output density matrix can be written in the form $\rho_{ab}^{out} = \mathcal{A}_{ab}^{local}~ \rho_{ab}^{0} ~\mathcal{A}_{ab}^{local\dag}$. Hence, the entanglement between the two modes of the output is encoded in the density matrix $\rho_{ab}^0$ since the operation $\mathcal{A}_{ab}^{local}$ only acts locally on the two modes.\\

\noindent\emph{Two-Mode Thermal State}: The
two-mode GSP operated thermal input state can be calculated using an approach similar to that for the two-mode input coherent state. The two modes are operated separately and then interact via a 50:50 beam-splitter [figure \ref{fig1}(b), bottom row].

The single-mode GSP operated thermal state can be written in a form similar to relations (\ref{9}), (\ref{10}). The action of the beam-splitter can be represented as the following:
\begin{eqnarray}
\rho_{ab}^{out} &=& \hat{\mathcal{B}}_{ab} \left(\rho_a^{'~th} \otimes \rho_b^{'~th} \right) \hat{\mathcal{B}}_{ab}^\dag \nonumber\\
&=& \hat{\mathcal{B}}_{ab} \left(p_1^a p_1^b |00\rangle\langle 00| + p_1^a p_2^b |01\rangle\langle 01| + p_2^a p_1^b |10\rangle\langle 10|\right. \nonumber\\
&& + \left. p_2^a p_2^b |11\rangle\langle 11|\right)\hat{\mathcal{B}}_{ab}^\dag ,
\label{17}
\end{eqnarray}
where $p_{n+1}^x = \left(\frac{M_x^{-1}}{1+\bar{n}_x}\right)\left( \frac{\bar{n}_x}{1+\bar{n}_x}\right) ^n \left( s^x + n \left( s^x +t^x \right)\right)^2$ corresponds to the two input thermal modes $x = a, b$. We have used the low average photon number approximation and retained terms up to $n <$ 2. Since the most of the terms in the GSP operated density matrix are product of the probabilities from the single-mode GSP operated thermal state, the smaller probabilities can be neglected for low average fields.

The two-mode output state generated, using (\ref{17}), is of the form
\begin{eqnarray}
\rho_{ab}^{out} &=& p_1^a p_1^b |00\rangle\langle 00|+ p_1^a p_2^b~ \hat{\mathcal{B}}_{ab}\hat{b}^\dag\hat{\mathcal{B}}_{ab}^\dag|00\rangle\langle 00|\hat{\mathcal{B}}_{ab} \hat{b}~\hat{\mathcal{B}}_{ab}^\dag \nonumber\\
&& + p_2^a p_1^b~ \hat{\mathcal{B}}_{ab}\hat{a}^\dag\hat{\mathcal{B}}_{ab}^\dag|00\rangle\langle 00|\hat{\mathcal{B}}_{ab} \hat{a}~\hat{\mathcal{B}}_{ab}^\dag \nonumber\\
&& + p_2^a p_2^b~ \hat{\mathcal{B}}_{ab}\hat{a}^\dag\hat{b}^\dag\hat{\mathcal{B}}_{ab}^\dag|00\rangle\langle 00|\hat{\mathcal{B}}_{ab} \hat{a}\hat{b}~\hat{\mathcal{B}}_{ab}^\dag .
\label{18}
\end{eqnarray}
Expanding the above relation (\ref{18}) by applying the beam-splitter operation, we obtain
\begin{eqnarray}
\rho_{ab}^{out} &=& p_1^a p_1^b |00\rangle\langle 00|+ p_1^a p_2^b~ \left(\frac{\hat{b}^\dag+\hat{a}^\dag}{\sqrt{2}}\right)|00\rangle\langle 00|\left(\frac{\hat{b}+\hat{a}}{\sqrt{2}}\right) \nonumber\\
&& + p_2^a p_1^b\left( \frac{\hat{a}^\dag-\hat{b}^\dag}{\sqrt{2}}\right)|00\rangle\langle 00|\left( \frac{\hat{a}-\hat{b}}{\sqrt{2}} \right) \nonumber\\
&& + p_2^a p_2^b \left( \frac{\hat{a}^\dag-\hat{b}^\dag}{\sqrt{2}}\right) \left( \frac{\hat{b}^\dag+\hat{a}^\dag}{\sqrt{2}}\right) |00\rangle\langle 00| \nonumber\\
&& \times \left( \frac{\hat{a}-\hat{b}}{\sqrt{2}}\right)\left(\frac{\hat{b}+\hat{a}}{\sqrt{2}}\right) .
\label{19}
\end{eqnarray}
On solving, we get
\begin{eqnarray}
\rho_{ab}^{out} &=& q'_1 |00\rangle\langle 00| + q'_2 \left(|01\rangle\langle 01|+|10\rangle\langle 10|\right) \nonumber\\
&& - q'_3 \left(|01\rangle\langle 10|+|10\rangle\langle 01|\right)+q'_4\left(|20 \rangle\langle 20|\right. \nonumber\\
&& + \left. |02 \rangle \langle 02| - |20 \rangle \langle 02| - |02 \rangle \langle 20| \right) ,
\label{20}
\end{eqnarray}
where $q'_1=p_1^a p_1^b$, $q'_2=\left( p_1^a p_2^b+p_2^a p_1^b\right)/2$, $q'_3=\left( p_2^a p_1^b-p_1^a p_2^b\right)/2$, and $q'_4=\left( p_2^a p_2^b\right)/2$.\\

The nonclassical properties of the two-mode output GSP operated thermal state can be obtained using the density matrix ($\rho_{ab}^{out}$) in relation (\ref{20}). \\

\noindent\emph{Two-Mode Squeezed State}: The two-mode squeezed state is a well-known Gaussian state that can be experimentally prepared using any pure Gaussian state upon unitary Gaussian operations. The initial squeezed state is an entangled two-mode state and hence it is not necessary to interact the modes via any active linear optical media as in the case of coherent or thermal input states. The two-mode squeezed state is an infinite dimensional superposition state of the two modes. The two-mode squeezed state can be written as,
\begin{eqnarray}
|\psi\rangle_{ab} &=& \exp[r~(\hat{a}^{\dag}\hat{b}^\dag-\hat{a}\hat{b})]|0\rangle_a|0\rangle_b \nonumber\\
&=& \sqrt{1-\lambda^2}~ \sum_{n=0}^\infty \lambda^n |n\rangle_a |n\rangle_b ,
\label{21}
\end{eqnarray}
where $r$ is the squeezing parameter and $\lambda$ = $\tanh r$. Applying the GSP operation on the two-mode squeezed state [figure \ref{fig1}(c), bottom row], we obtain
\begin{eqnarray}
|\psi'\rangle_{ab} &=& \sqrt{\frac{1}{N}} ~(s_1 \hat{a} \hat{a}^\dag + t_1 \hat{a}^\dag \hat{a})(s_2 \hat{b} \hat{b}^\dag + t_2 \hat{b}^\dag \hat{b})\nonumber \\
&& \times \sqrt{1-\lambda^2} \sum_{n=0}^\infty \lambda^n |n\rangle_a |n\rangle_b \nonumber\\
&=& \sqrt{\frac{1-\lambda^2}{N}} \sum_{n=0}^\infty \lambda^n \left[ s_1 s_2 + n(s_1(s_2+t_2) \right. \nonumber\\
&& + s_2(s_1+t_1) + n^2(s_1+t_1)(s_2+t_2)) \nonumber\\
&& \left. \times |n\rangle_a |n\rangle_b \right] ,
\label{22}
\end{eqnarray}
where $N$ is the normalization constant. The infinite dimensional two-mode squeezed state can be reduced by considering the low squeezing region. For $\lambda \ll$ 1, the GSP operated squeezed state reduces to
\begin{eqnarray}
|\psi'\rangle_{ab} &=& \sqrt{\frac{1-\lambda^2}{N}} \left[ s_1 s_2 |0\rangle_a|0\rangle_b + \lambda [s_1s_2+s_1(s_2+t_2) \right.\nonumber\\
&+& \left. s_2(s_1+t_1)+(s_1+t_1)(s_2+t_2)]|1\rangle_a|1\rangle_b \right] .
\label{23}
\end{eqnarray}
The GSP operated two-mode squeezed state in relation (\ref{23}) gives us the two-mode output pure state density matrix ($\rho_{ab}^{out}=|\psi'\rangle \langle \psi'|_{ab}$) that can be used to study its nonclassical properties.

\begin{figure}[ht]
\begin{center}
\subfigure[]{\includegraphics[width=5.2cm]{Fig2a_Ent_1coh.eps}}\hspace{0.1cm}
\subfigure[]{\includegraphics[width=5.2cm]{Fig2b_Ent_1thm.eps}}\hspace{0.1cm}
\subfigure[]{\includegraphics[width=5.2cm]{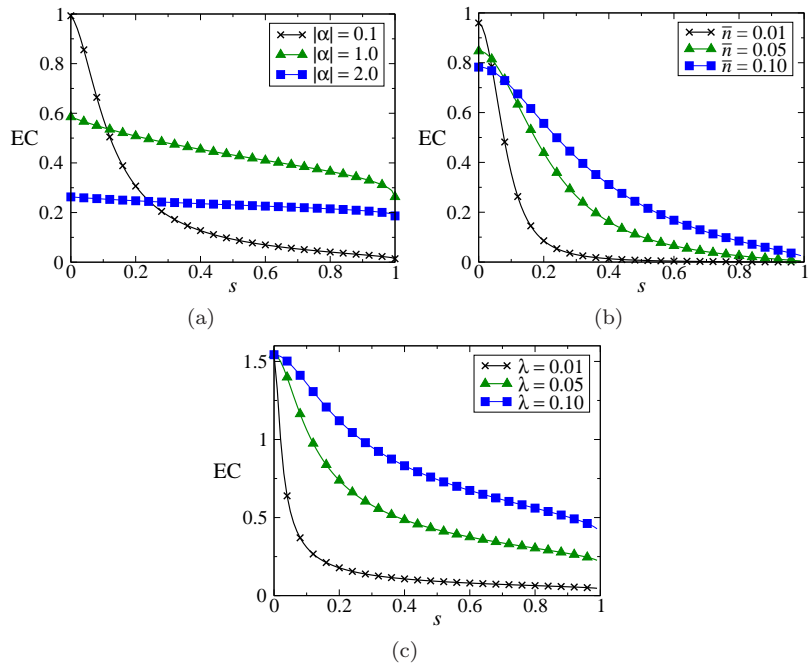}}
\caption{Behaviour of the entanglement capacity (EC), for two-mode output states generated from GSP operated single-mode input states interacting with a vacuum mode. EC is shown as a function of the GSP operator parameter $s$ for (a) different amplitudes ($|\alpha|$ = 0.1, 1.0, 2.0) of a single-mode coherent input state, (b) different values of average photon number ($\bar{n}$ = 0.01, 0.05, 0.10) for a thermal input state, and (c) different values of $\lambda (\equiv \tanh z$ = 0.01, 0.05, 0.10), where $z$ is the squeezing parameter of a squeezed input state.}
\end{center}
\label{ent1}
\end{figure}

\begin{figure}[htb]
\begin{center}
\subfigure[]{\includegraphics[width=3.4cm,angle=-90]{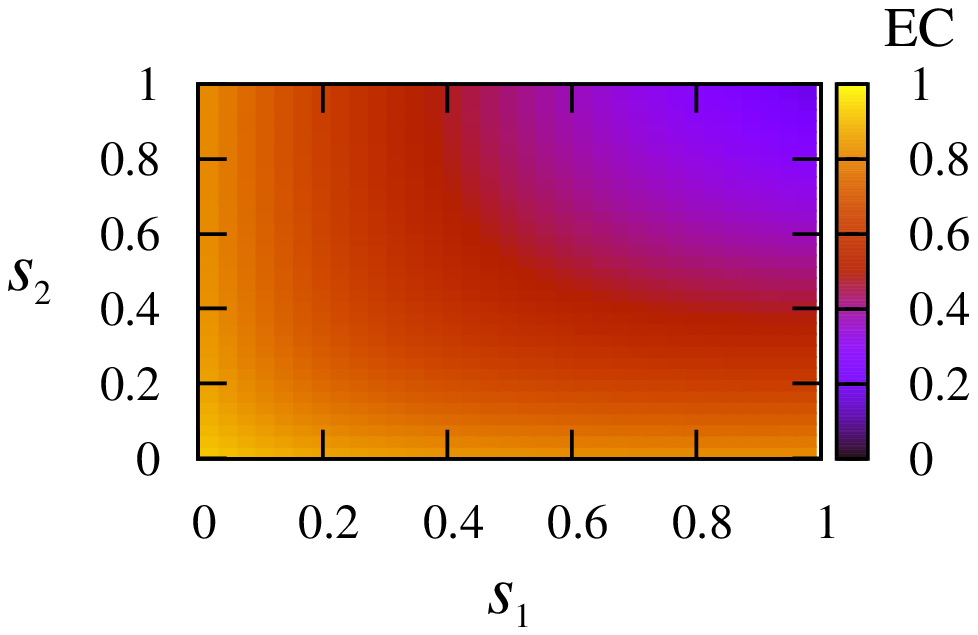}}\hspace{0.1cm}
\subfigure[]{\includegraphics[width=3.4cm,angle=-90]{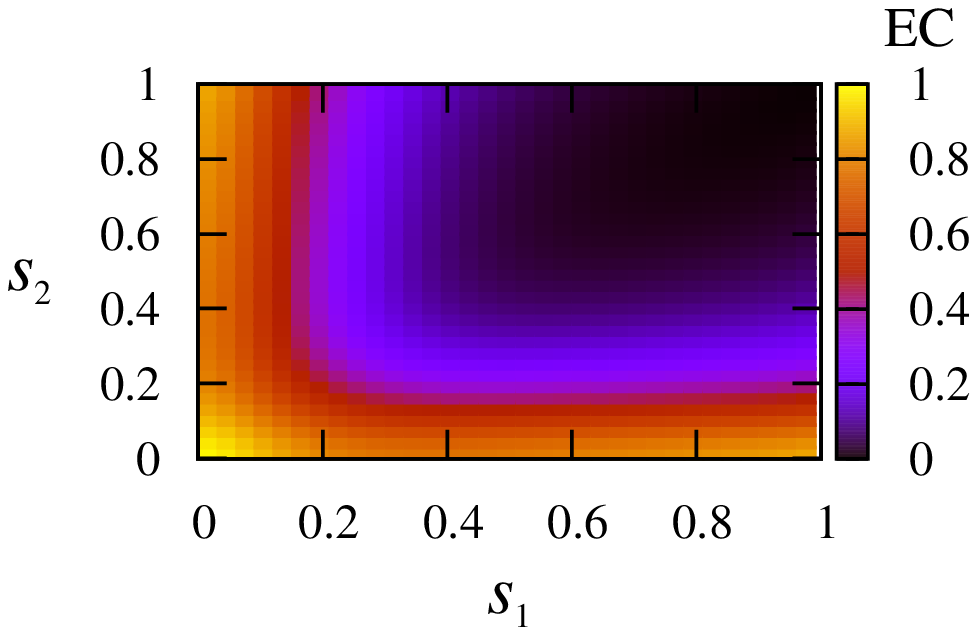}}\hspace{0.1cm}
\subfigure[]{\includegraphics[width=3.4cm,angle=-90]{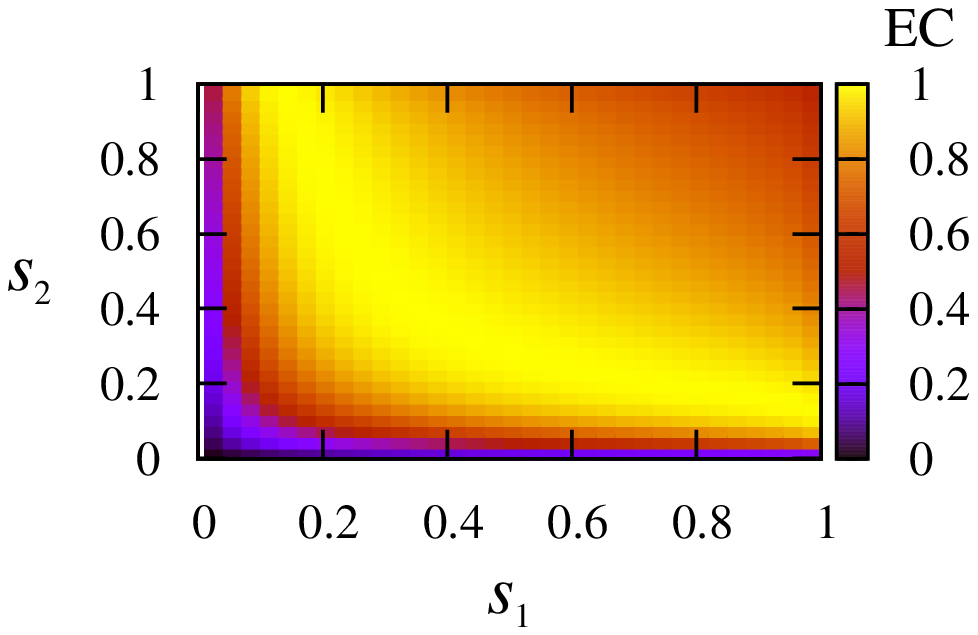}}\hspace{0.1cm}
\caption{Behaviour of the entanglement capacity (EC) for two-mode output states generated from GSP operated two-mode input states. The variation of EC is shown with respect to the two-mode GSP operator parameters $s_1$ and $s_2$ for (a) coherent state amplitudes $|\alpha|$ = $|\beta|$ = 0.5, for a two-mode coherent input state, (b) fixed average photon number, $\bar{n}_a$ = $\bar{n}_b$ = 0.05, for a two-mode thermal input state, and (c) for a fixed value of $\lambda ~(\equiv \tanh r)$ = 0.05, where $r$ is the squeezing parameter of the two-mode squeezed input state.}
\end{center}
\label{ent2}
\end{figure}

\section{Entanglement properties}
\label{ss5}

As discussed in sections \ref{ss2} and \ref{ss3}, using a low average field-intensity regime, the infinite-dimensional output optical state can be mapped to a discrete-level optical state. The entanglement properties of the single- and two-mode GSP operated input states is characterized using the entanglement capacity discussed in section \ref{ent}, where logarithmic negativity (\ref{logneg}) is used as the measure of two-mode entanglement.
The parameters that control the EC of the final two-mode states are the parameter $s$ of the GSP operator and the initial field parameters of the input states, such as amplitude of the coherent state, average photon number of the thermal state and squeezing parameter of the squeezed state.\\

\noindent\emph{Single-mode states}:
The variation of EC with GSP operation parameter $s$ and field parameters for the single-mode input states is given in figure \ref{ent1}. In figure \ref{ent1}(a), we observe that high EC is observed for lower values of $s$ and the amplitude ($|\alpha|$) of the coherent state input. The operation generates maximal entanglement at $s \approx 0$. For greater $|\alpha|$, the states have lower values of EC but are not very sensitive to variation in $s$.
We also observe that for the truncated input thermal state [figure \ref{ent1}(b)], the EC is maximum at $s \approx 0$ and steadily decreases with increase in $s$. The maximum EC corresponds to the lowest average photon number ($\bar{n} = 0.01$). This is due to the fact that the classicality of coherent and thermal states increases with increase in average photon number. Figure \ref{ent1}(c) gives us the variation of EC with $\lambda (=\tanh z)$, where $z$ is the squeezing parameter of the input single-mode squeezed state. The maximum EC is obtained at $s > 0$, and the EC decreases with increasing $s$. The state is seen to be more entangled for higher values of $\lambda$ for all values of $s$ away from $s$ = 0. Hence, the squeezed output with high $\lambda$ is nonclassical for all ranges of the GSP operation.\\

\noindent\emph{Two-mode states}: The EC of the GSP operated two-mode coherent, thermal and squeezed input states is shown in figure \ref{ent2}. The variation of the EC is with respect to the two-mode GSP operation parameters $s_1$ and $s_2$ for fixed values of field parameters such as amplitude, average photon number and squeezing parameter for coherent, thermal and squeezed states, respectively. For coherent and thermal fields, we consider inputs with equal field parameters for both the input modes.

Figure \ref{ent2}(a) gives us the variation of EC with the two-mode GSP operator parameters $s_1$ and $s_2$ for the two-mode coherent state input with equal amplitudes, $|\alpha|$ = $|\beta|$ = 0.5. We observe that low values of either of the mode operators $s_1$ and $s_2$ are sufficient for obtaining entangled states. The only region that produces low entanglement is the region corresponding to simultaneous high values of both $s_1$ and $s_2$ marked by the dark region in the colour scheme of figure \ref{ent2}(a). In contrast, the plot for the EC of two-mode thermal state input in figure \ref{ent2}(b), with fixed photon number averages $\bar{n}_a$ = $\bar{n}_b$ = 0.05, shows that regions of relatively high entanglement is restricted to a small region along the two axes that correspond to the small values of $s_1$ ($\ll 1$) or $s_2$ ($\ll 1$). For other values of $s_1$ and $s_2$, the output state has very low entanglement. The behaviour of EC for the two-mode squeezed state ($\lambda$ = 0.05) is completely opposite of that of the two-mode thermal state. Figure \ref{ent2}(c) shows that the EC remains high at most values of $s_1$ and $s_2$, dropping in the region of $s_1, s_2 \approx 0$. The two-mode input squeezed state, unlike the two-mode coherent and thermal state, is entangled prior to the GSP operation. However, the GSP operation enhances the amount of entanglement \cite{refyang} over the range of parameters $s_1$ and $s_2$.

Hence, it is observed that the GSP operation introduces nonclassicality in single- and two-mode coherent, thermal and squeezed states, which in turn can be characterized by high entanglement capacity. This clearly shows that GSP operation, in the low field-intensity limit and in suitable parameter regimes, can be used to generate highly entangled two-mode optical states, for all the three inputs considered in our investigation. To investigate the utility of the generated nonclassical and entangled two-mode states, under GSP operations, we use these states as quantum resources in specific quantum information protocols. In particular, we investigate the advantage obtained by using these two-mode entangled states as quantum channels in continuous variable quantum teleportation.

\section{Average Fidelity of Continuous Variable Teleportation}
\label{ss6}

Quantum teleportation was originally devised as a quantum information protocol that involved sending an unknown qubit, from one party to another, across an entangled EPR channel shared between the two parties \cite{Qtele}. In the CV regime, the earliest example of teleportation was formulated for a one-dimensional phase-space particle \cite{vaid}. A more practical version of the CV quantum teleportation was devised by Braunstein and Kimble \cite{BK}, who proposed the teleportation of quadrature components of an electromagnetic field using finite degrees of correlation. Such CV quantum teleportation protocol is experimentally realizable \cite{furu}.

In this section, we investigate the resourcefulness of the two-mode entangled output states generated in sections \ref{ss2} and \ref{ss3} under GSP operations, in CV quantum teleportation. In particular, we explore whether the nonclassicality generated under the GSP operations is useful in providing a quantum advantage in specific information protocol such as teleportation. The success of the CV quantum teleportation, under the BK protocol, is indicated by the average fidelity ($F$) between the final output and the initial input state being teleported. A genuine quantum advantage under the GSP operation will be observed if it can enhance the average fidelity of the CV teleportation protocol.

Under the BK protocol, perfect teleportation fidelity ($F = 1$) is obtained with an infinitely nonclassical channel such as the ideal EPR entangled state \cite{BK}. For non-ideal channels, the average fidelity is less than 1 and the maximum value attainable using a classical channel is $F_{class} = \frac{1}{2}$ \cite{class}. This is the maximum permissible fidelity without the use of any entangled quantum channels. An important figure of merit in CV teleportation is the no-cloning limit, which is equal to $\frac{2}{3}$ \cite{clone}. To ensure that the teleported state is the best copy of the state remaining after the protocol and the nonclassical features of the input state have been teleported \cite{furusawa}, the average fidelity must be greater than the no-cloning limit. Hence, one can define a fidelity of cloning $F_{clone} = \frac{2}{3}$, which is an important benchmark in the success of the protocol.

In the CV quantum teleportation formulated by BK, the two-mode quantum channel and the input state to be teleported are described by a joint Wigner function $W(\gamma,\xi,\eta)=W_{in}(\gamma) \otimes W_{ch}(\xi,\eta)$. The Wigner function description can be more clearly stated in terms of the symmetrically ordered characteristic function \cite{Scully, Barnett}. The average fidelity of the quantum teleportation is \cite{fidel}
\begin{equation}
F=\frac{1}{\pi}\int d^2\gamma ~\chi_{in}(\gamma)\chi_{out}(-\gamma) ,
\label{24}
\end{equation}
where (from \cite{41})
\begin{equation}
\chi_{out}(\gamma)=\chi_{in}(\gamma)\chi_{ch}(\gamma^*,\gamma),
\label{25}
\end{equation}
$\chi_{in}$ and $\chi_{ch}$ being the characteristic functions of the input state to be teleported and the two-mode quantum channel, respectively.
Hence, an important part of the calculation of the average fidelity of CV teleportation is to obtain the characteristic function of the state to be teleported and the two-mode quantum channel. In our work, we use the GSP operated, two-mode entangled states generated in sections \ref{ss2} and \ref{ss3}, as the quantum channel in the teleportation protocol. In the following, we show how the characteristic functions for the GSP operated single- and two-mode coherent, thermal and squeezed states can be derived in order to calculate the average fidelity of teleportation. \\

\begin{figure}[htb]
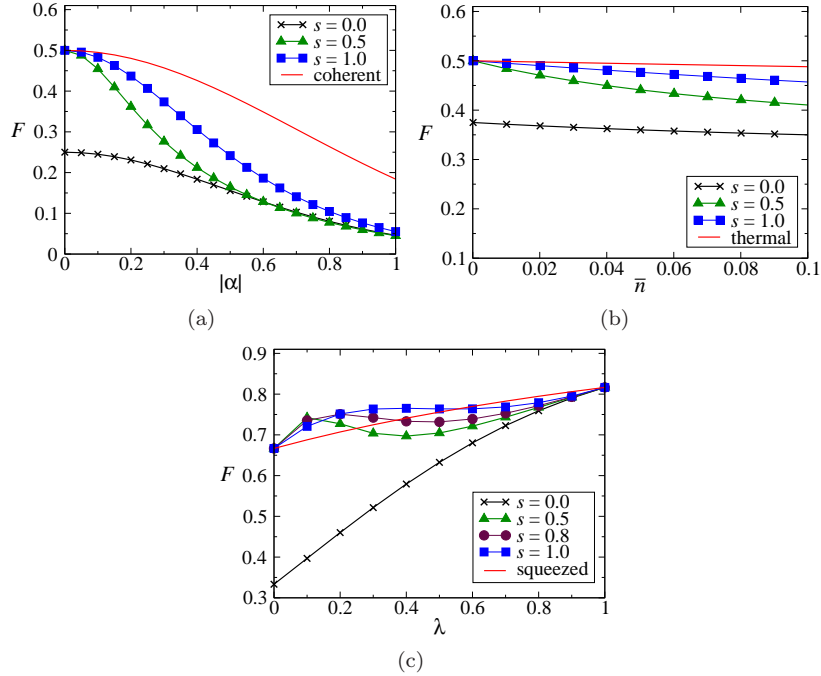

\begin{center}
\subfigure[]{\includegraphics[width=5.2cm]{Fig4a_fidcoh_1coh.eps}}\hspace{0.1cm}
\subfigure[]{\includegraphics[width=5.35cm]{Fig4b_fidcoh_1thm.eps}}\hspace{0.1cm}
\subfigure[]{\includegraphics[width=5.2cm]{Fig4c_fidcoh_1sqz.eps}}
\caption{Behaviour of the average fidelity ($F$) of teleporting a single-mode coherent state (of any amplitude) using two-mode output states generated from GSP operated single-mode input states interacting with a vacuum mode. $F$ is shown as a function of (a) the amplitude ($|\alpha|$) of a coherent input state, (b) the average photon number ($\bar{n}$) of a thermal input state, and (c) $\lambda ~(=\tanh z)$, where $z$ is the squeezing parameter of a squeezed input state. The plots are for different values of the GSP operator parameter: $s$ = 0 (black crosses), $s$ = 0.5 (green triangles), $s$ = 0.8 (brown circles) [shown only in (c)], $s$ = 1.0 (blue squares), and unoperated states (red continuous line).
}
\end{center}
\label{fid1}
\end{figure}

\noindent\emph{Single-mode states}: We have considered three different single-mode input states for the GSP operation, namely, coherent, thermal and squeezed states. These input states undergo GSP operation and then interact via a symmetric beam-splitter with a vacuum mode to generate a two-mode bipartite entangled output state, as shown in section \ref{ss2}. The generated two-mode nonclassical states are used as a quantum channel in CV teleportation.

For a single-mode GSP operated coherent state coupled to a vacuum mode, the final two-mode nonclassical state can be written as
\begin{equation}
|\phi\rangle_{ab}^{out}=\hat{B}_{ab}(s+(s+t)\hat{a}^\dag\hat{a})|\alpha,0\rangle .
\label{27}
\end{equation}
The characteristic function of the state (\ref{27}) is given by
\begin{eqnarray}
\chi_{ch}(\xi,\eta) &=& N^{-1}\big(s^2 + (s+t)^2 |\alpha|^2\{1+(\alpha + X)(\alpha^*-X^*)\} \nonumber \\
&& + s(s+t)\{\alpha (\alpha^*-X^*)+\alpha^*(\alpha+X)\}\big) \nonumber\\
&& \times \exp\left[ \frac{1}{2}(|X|^2+|Y|^2)\right] \exp(\alpha^*X-\alpha X) ,
\label{28}
\end{eqnarray}
where $X=(\xi-\eta)/\sqrt{2}$ and $Y=(\xi+\eta)/\sqrt{2}$. $N$ is the normalization constant given by $[s^2+(s+t)(3s+t)|\alpha|^2+(s+t)^2|\alpha|^4]$.
The average fidelity can be calculated using relations (\ref{24}) and (\ref{25}) provided the characteristic function of the state to be teleported is also known.

For the GSP operated single-mode thermal state, the characteristic function can be calculated using the density matrix expression for the truncated GSP operated two-mode density matrix given by relation (\ref{13}). The characteristic function of any two-mode state with a density matrix $\rho$ is given by
\begin{equation}
\chi(\xi,\eta)=\sum_{n',n,m',m=0}^\infty \rho_{nm,n'm'}\langle n m|\hat{D}_a(\xi)\hat{D}_b(\eta)|n' m'\rangle ,
\label{29}
\end{equation}
where $\hat{D}_a$ is the displacement operator acting on mode $a$. The detailed derivation of the characteristic function for any two-mode density matrix in the number state basis is shown in \ref{app1}. The exact analytic expression for the characteristic function of the GSP operated single-mode thermal state can be calculated using (\ref{29}). The average fidelity is then calculated using expressions (\ref{24}) and (\ref{25}).

The characteristic function of a single-mode GSP operated squeezed state interacting with a beam-splitter can be written as
\begin{eqnarray}
\chi(\xi,\eta) &=& N^{-1}\big(s^2 + s(s+t)\{2p_0 - p_1(X'^{*2}+X'^2)\}\nonumber\\
&& + s^2(s+t)^2\{p_0^2-p_0p_1(X'^{*2}+X'^2)+p_1^2 \nonumber\\
&& \times (1-2|X'|^2+\frac{|X'|^4}{2}\}\big) \nonumber\\
&& \times \exp\left[ -\frac{1}{2}(|X'|^2+|Y|^2)\right] ,
\label{sqz1ch}
\end{eqnarray}
where $X=(\xi-\eta)\sqrt{2}$ and $Y=(\xi+\eta)\sqrt{2}$. $X'=X\cosh z-Y^*\sinh z$. $p_0=\sinh ^2 z$ and $p_1=\sinh z \cosh z$. The normalization constant is $N=s^2+ 2s(s+t)p_1+(s+t)^2(p_1^2+2p_2^2)$. The symbols $s$, $t$ and $z$ have the usual meanings as in section \ref{ss2}. The average fidelity can be calculated using the characteristic function (\ref{sqz1ch}).\\

\noindent\emph{Two-mode states}: Now we consider three two-mode input states in our analysis. The first two are the two-mode GSP operated coherent and thermal input states interacting via a symmetric beam splitter to generate a bipartite entangled two-mode nonclassical output. The third is the GSP operated two-mode squeezed state. The generation of the nonclassical GSP operated states have been shown in section \ref{ss3}.

The calculation of the characteristic function for the two-mode GSP operated input is more complicated than for the case of single-mode input interacting with vacuum mode. For a two-mode coherent input state, the derivation of the characteristic function of the GSP operated final bipartite state (\ref{16}) is given in \ref{app2}. The average fidelity of the teleportation of a state across the nonclassical two-mode GSP operated coherent input can be calculated using relations (\ref{24}) and (\ref{25}).
The characteristic function of the two-mode GSP operated thermal state can be derived using the relation (\ref{29}). The complete steps for the derivation of the characteristic function is given in \ref{app1}. The density matrix elements in the number state representation required for calculating the characteristic function in (\ref{29}) are given by the GSP operated two-mode output state given by relation (\ref{20}).
The characteristic function of the final bipartite state (\ref{22}) obtained by GSP operated two-mode squeezed state input is given in \ref{app3}. Note that the truncated state (valid for low field intensity) is needed for the calculation of characteristic function only in the case of the thermal input state.

The characteristic function of the nonclassical output states of the GSP operated single and two-mode input is used for calculating the average fidelity in teleporting a single-mode state, using these GSP generated states as quantum channels. In the following subsections, we consider the teleportation of two single-mode states: (a) single-mode coherent state, and (b) single-mode squeezed state.

\subsection{Teleporting a single-mode coherent state}
\label{ss6a}

We consider the teleportation of a single-mode coherent state using the GSP operated two-mode states as a quantum teleportation channel, generated using single- and two-mode coherent, thermal and squeezed state, as shown in the previous sections. The teleportation is conducted and optimized using the BK protocol. The symmetrically ordered characteristic function of the input coherent state $|\alpha\rangle$ is
\begin{equation}
\chi^{coh}(\gamma)=\exp\left[ {-\frac{1}{2}|\gamma|^2}\right] \exp({\alpha^*\gamma-\alpha\gamma^*}).
\label{26}
\end{equation}
The suitability of a two-mode entangled quantum channel to successfully transport a single-mode state under the BK teleportation protocol can be measured by calculating the average fidelity using Eqs.\,(\ref{24}) and (\ref{25}).
The fidelity of teleportation of a coherent state input under the BK protocol is independent of the amplitude of the state to be teleported.

\begin{figure}
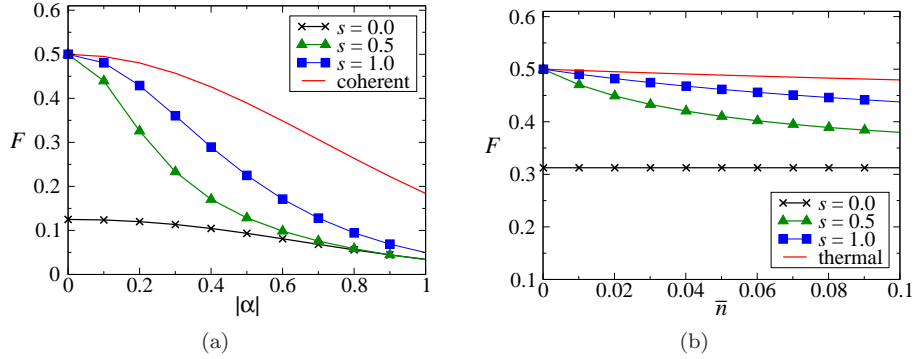

\begin{center}
\subfigure[]{\includegraphics[width=5.6cm]{Fig5a_fidcoh_2coh.eps}}\hspace{0.6cm}
\subfigure[]{\includegraphics[width=5.7cm]{Fig5b_fidcoh_2thm.eps}}
\caption{Behaviour of the average fidelity ($F$) of teleporting a single-mode coherent state (of any amplitude) using a two-mode output generated from GSP operated two-mode input states. $F$ is shown as a function of (a) the amplitude ($|\alpha|$) of a coherent input state, and (b) the average photon number ($\bar{n}$) of a thermal input state. The plots are for different symmetric values of the GSP operator parameters, $s_1 = s_2 = s$: $s$ = 0 (black crosses), $s$ = 0.5 (green triangles), $s$ = 1.0 (blue squares), and unoperated states (red continuous line).}
\end{center}
\label{fid2cohthm}
\end{figure}

Let us consider the average fidelity of teleporting a single-mode coherent state via a two-mode output state generated from GSP operations on single-mode states. The characteristic function of the GSP operated single-mode coherent, thermal and squeezed input interacting with vacuum is given by Eqs.\,(\ref{28}), (\ref{29}) and (\ref{sqz1ch}), respectively. Figure \ref{fid1} shows the variation of the average fidelity ($F$) with changing field parameters of the quantum channel at different values of the GSP operation parameter $s$. We observe in figure \ref{fid1}(a), that for the GSP operated single-mode coherent state coupled with vacuum channel the maximum average fidelity is, $F_{max} \approx$ 0.5. This value is the classical upper limit of teleportation, $F_{class}$, with an unentangled source and is well below the threshold of cloning fidelity. The maximum fidelity is achieved at very low values of the channel field amplitude $|\alpha|$ and is independent of the GSP operator parameter $s$. 
Interestingly, the decaying $F$ with increasing $|\alpha|$ for the GSP operated state is always bounded above by that for the non-operated two-mode state. Hence, the GSP operations reduce the teleportation fidelity under optimizations in the BK protocol \footnote{Though the average fidelity is below 0.5 under specific optimizations in the BK protocol, the local degrees of the channel can be optimized outside the protocol to obtain an average fidelity of 0.5 without using any extra quantum resource. This is due to the fact that no entanglement is needed to achieve the classical fidelity.}.
Figure \ref{fid1}(b) gives us the fidelity for the GSP operated single-mode thermal state coupled with vacuum channel for varying average photon number. The maximum fidelity, $F_{max} \approx$ 0.5, which, again, is the classical threshold. Hence, the GSP operated single-mode coherent and thermal channels are not resourceful for quantum teleportation, even though they are entangled, as shown in section \ref{ss5}. Similarly, the fidelity of the GSP operated thermal state is always bounded above by the fidelity of the non-operated thermal state. Hence, the operation reduces the average fidelity in GSP operated thermal channels.

The enhancement of average fidelity can be observed for the quantum channel generated from the GSP operated single-mode squeezed state. From figure \ref{fid1}(c), we can observe that a high fidelity, $F_{max} \approx$ 0.816, can be achieved for highly squeezed states. This value breaches the threshold for quantum cloning. In the low squeezing regime, GSP operation can highly enhance the average fidelity. For low squeezing parameters, $\lambda \le$ 0.2, the GSP operated states at $s \ge$ 0.5 offer enhanced fidelity values, $F \ge$ 0.74, which is again higher than the cloning fidelity, $\frac{2}{3}$ and also higher than the non-operated squeezed channel.

Hence, one observes that the nonclassical states generated using single-mode coherent and thermal states do not offer any quantum advantage in CV quantum teleportation, even though they are discrete level two-mode entangled. However, nonclassical states formed by GSP operation on initially nonclassical squeezed states offer a distinct advantage in enhancing average fidelity.

\begin{figure}[h]
\begin{center}
\epsfig{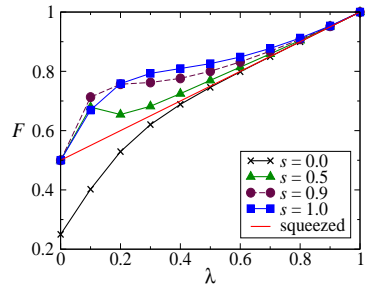}
\caption{Behaviour of the average fidelity ($F$) of teleporting a single-mode coherent state (of any amplitude) using a two-mode output generated from a GSP operated two-mode input squeezed state. $F$ is shown as a function of $\lambda ~(=\tanh r)$, where $r$ is the squeezing parameter of the squeezed state input. The plots are for different symmetric values of the GSP operator parameters, $s_1 = s_2 = s$: $s$ = 0 (black crosses), $s$ = 0.5 (green triangles), $s$ = 0.9 (brown circles), $s$ = 1.0 (blue squares), and unoperated states (red continuous line).}
\label{fid2sqz}
\end{center}
\end{figure}

Let us now consider the average fidelity of teleporting a single-mode coherent state via a bipartite two-mode output generated from a GSP operated two-mode input state. 
The average fidelity ($F$) for the two-mode GSP operated coherent and thermal state channel can be observed in figure \ref{fid2cohthm} as a function of the field parameters $|\alpha|$ and $\bar{n}$. We have considered symmetric field parameters for the two modes. The behaviour of the average fidelity for the two-mode case is similar to the results for the single-mode fidelity. For the two-mode GSP operated coherent and thermal input the maximum average fidelity is again $F_{max} \approx$ 0.5. The two-mode GSP operated coherent and thermal state has a maximum fidelity attainable by a classical channel. However, the average fidelity is always bounded by the non-operated state. 

Figure \ref{fid2sqz} gives us the average fidelity ($F$) for a two-mode GSP operated squeezed state channel with varying field parameter $\lambda$. $\lambda$ = $\tanh r$, where $r$ is the two-mode squeezing parameter. $F$ is calculated for different values of the GSP operator parameter $s$. For $s=$ 0 and $s=$ 1, the fidelity, $F$, has an analytical solution:
\begin{eqnarray}
F_{s=0} &=& \frac{(1+\lambda^5)((1 + \lambda +  \lambda^2 )}{4 (1 + 11 \lambda^2 + 11 \lambda^4 + \lambda^6)}, \nonumber\\
F_{s=1} &=& \frac{(1+\lambda^5)((2 - 2 \lambda + 5 \lambda^2 - 3 \lambda^3 + \lambda^4)}{4 (1 + 11 \lambda^2 + 11 \lambda^4 + \lambda^6)} . \nonumber
\end{eqnarray}
The maximum fidelity, $F_{max}$ = 1 is achieved for all values of $s$ at high squeezing ($\lambda$ = 1). In the low-squeezing regime, for $\lambda \le$ 0.2, the highest fidelity, $F \approx$ 0.76 is achieved at $s$ = 0.9. At slightly higher squeezing, $0.2 \le \lambda \le 0.8$, the highest fidelity, $F \approx$ 0.91 is achieved at $s$ = 1.0.

\begin{figure}[h]
\begin{center}
\epsfig{figure=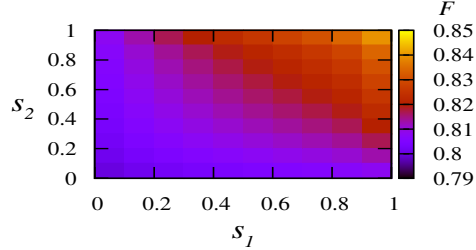, height=.3\textheight, width=.25\textwidth, angle=-90}
\caption{Behaviour of the average fidelity ($F$) of teleporting a single-mode coherent state (of any amplitude) using a two-mode output generated from a GSP operated two-mode input squeezed state. $F$ is shown as a function of the two-mode GSP operator parameters $s_1$ and $s_2$, with $\lambda ~(=\tanh r)$ fixed at 0.6, where $r$ is the squeezing parameter of the two-mode squeezed input state.}
\label{fid2sqz3d}
\end{center}
\end{figure}

Figure \ref{fid2sqz3d} shows the variation of average fidelity $F$ with the two-mode GSP operator parameters $s_1$ and $s_2$ applied on a two-mode squeezed state, with $\lambda$ = 0.6. The average fidelity is independent of the amplitude of the single-mode coherent state to be teleported. $F$ varies over a small range of values (0.79--0.85) for different values of $s_1$ and $s_2$, with the higher range corresponding to higher values of the GSP operator parameters. Hence, for a fixed squeezing, $\lambda$, the fidelity is not very sensitive to the operator parameters.

Hence, we once again observe that GSP operated coherent and thermal states offer no advantage in CV quantum teleportation, in contrast to the performance of the GSP operated squeezed states.

\subsection{Teleporting a single-mode squeezed state}
\label{ss6b}

We now consider the teleportation of single-mode squeezed vacuum state using the GSP operated two-mode output states discussed in section \ref{ss2} and \ref{ss3} as the teleportation channel. As for the single-mode coherent input, the teleportation is conducted using the BK protocol. The symmetrically ordered characteristic function of the input squeezed vacuum state, $\exp[(r'/2)(\hat{a}^2-\hat{a}^{\dag 2})]|0\rangle$ is
\begin{equation}
\chi_{sqz}(\gamma) = \exp\left[ -\frac{\cosh 2r'}{2}|\gamma|^2-\frac{\sinh 2r'}{4}(\gamma^2+\gamma^{*2})\right] .
\end{equation}

For the teleportation of the single-mode squeezed state, we consider the single- and two-mode GSP operated coherent and squeezed state quantum channels using the BK protocol. The average fidelity of the BK protocol is not independent of the squeezing parameter $r'$ of the squeezed state to be teleported. The fidelity for the coherent and thermal channel is very similar to the fidelity obtained for teleporting a single-mode coherent state.

\begin{figure}[t]
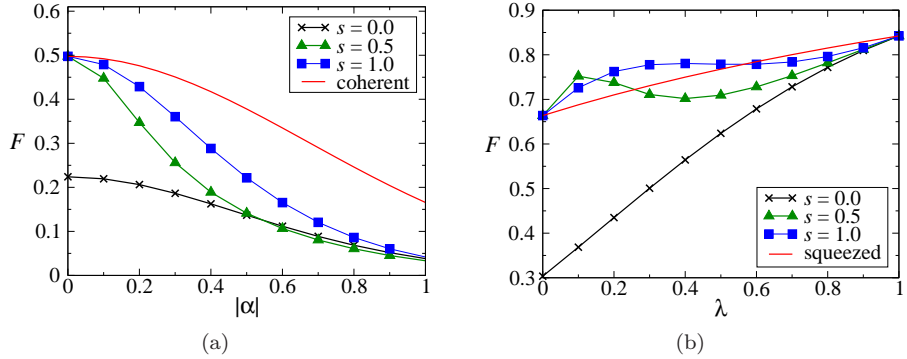

\begin{center}
\subfigure[]{\includegraphics[width=5.6cm]{Fig8a_fidsqz_1coh_r_0.1.eps}}\hspace{0.6cm}
\subfigure[]{\includegraphics[width=5.6cm]{Fig8b_fidsqz_1sqz_r_0.1.eps}}
\caption{Behaviour of the average fidelity ($F$) of teleporting a single-mode squeezed state with squeezing parameter, $r'$ = 0.1, using a two-mode output generated from GSP operated single-mode input states interacting with vacuum mode. $F$ is shown as a function of (a) the amplitude ($|\alpha|$) of a coherent input state, and (b) $\lambda ~(=\tanh z)$, where $z$ is the squeezing parameter of a squeezed input state. The plots are for different values of the GSP operator parameter: $s$ = 0 (black crosses), $s$ = 0.5 (green triangles), $s$ = 1.0 (blue squares), and unoperated states (red continuous line).}
\end{center}
\label{fidsqz1}
\end{figure}

Figure \ref{fidsqz1} shows the average fidelity ($F$) of teleporting a single-mode squeezed state, with squeezing parameter $r'$=0.1, using a GSP operated single-mode coherent and squeezed state input coupled with a vacuum mode. For a single-mode GSP operated coherent state, as shown in figure \ref{fidsqz1}(a), the maximum average fidelity is $F_{max} \approx$ 0.5, which is the classical threshold. The GSP operation again reduces the average fidelity of teleportation. Figure \ref{fidsqz1}(b) shows the average fidelity of teleportation using a single-mode GSP operated squeezed state channel. The maximum average fidelity is $F_{max}$ = 0.842, at $\lambda$ = 1, for highly squeezed channels. At low squeezing limits, the GSP operations enhance the average fidelity over the non-operated state. For $\lambda \le$ 0.2, the highest fidelity is 0.752 for $s \approx$ 0.5.

\begin{figure}
\begin{center}
\epsfig{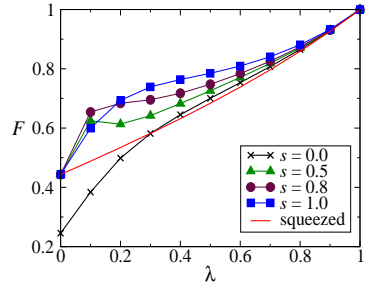}
\end{center}
\caption{Behaviour of the average fidelity ($F$) of teleporting a single-mode squeezed state with squeezing parameter, $r'$ =0.5, using a two-mode output generated from a GSP operated two-mode squeezed input state. $F$ is shown as a function of $\lambda ~(=\tanh r)$, where $r$ is the squeezing parameter of the input state. The plots are for different symmetric values of the GSP operator parameters, $s_1 = s_2 = s$: $s$ = 0 (black crosses), $s$ = 0.5 (green triangles), $s$ = 0.8 (brown circles), $s$ = 1.0 (blue squares), and unoperated states (red continuous line).}
\label{fidsqz2}
\end{figure}

Figure \ref{fidsqz2} shows the average fidelity ($F$) of teleporting a single-mode squeezed state, with squeezing parameter $r'$ = 0.5, using a GSP operated two-mode squeezed state input. Since $F$ is dependent on the squeezing parameter $r'$ of the state to be teleported, there is no analytical solutions for the fidelity. The maximum fidelity is $F_{max}$ = 1, achieved for highly squeezed channels with $\lambda \approx$ 1.0. At the mid-squeezing range, in the vicinity of $\lambda \approx$ 0.5, we observe that all GSP operated states have enhanced average fidelity over the non-operated state, with the highest fidelity, $F$ = 0.785, achieved at $s$ = 1.0. The behaviour of the average fidelity for a two-mode GSP operated coherent channel is similar to the single-mode coherent channel.

To summarize, we find that the GSP operation, while enhancing the teleportation fidelity for the nonclassical two-mode squeezed input state channel, reduces the fidelity for channels formed by GSP operated classical thermal and coherent input states, under the BK protocol. Hence, the nonclassicality generated under GSP operations does not directly lead to any genuine advantage in quantum information protocols for all input states. In particular, nonclassical states generated from classical inputs, such as coherent and thermal, perform no better in teleportation than classical channels. This is a significant result showing that the nonclassicality under non-Gaussian operations, such as the GSP, does not necessarily contribute to advantages in teleportation. Another significant ramification is for GSP operations on noisy quantum channels, which can be modelled using a linear characteristic function of a low intensity thermal and an entangled quantum channel \cite{noisy}. Using GSP operations, under the BK protocol, the average fidelity of teleportation under the quantum channel is enhanced while the noise is reduced.

To analyze the results obtained for CV teleportation using the GSP operated nonclassical channels, we consider the EPR correlations in these two-mode states. The nonclassicality of the state in terms of EPR correlations will give us a better insight into the performance of these channels in the teleportation protocol, and in particular, it will clarify why certain GSP operated nonclassical states do not provide any quantum advantage in teleportation.

\section{EPR correlations}
\label{sec-epr}

The success of the CV teleportation protocol is dependent on the correlations developed between the phase-quadrature components of the GSP operated two-mode output states. These correlations also known as EPR (Einstein-Podolosky-Rosen) correlations can be quantified in terms of the variance of the differences of the quadrature components of the two modes. The quadrature operators are defined as
\begin{eqnarray}
\hat{x}_1 &=& \frac{1}{\sqrt{2}}(\hat{a}+\hat{a}^\dag), ~\hat{p}_1 = \frac{1}{i\sqrt{2}}(\hat{a}-\hat{a}^\dag), \nonumber\\
\hat{x}_2 &=& \frac{1}{\sqrt{2}}(\hat{b}+\hat{b}^\dag), ~\hat{p}_2 = \frac{1}{i\sqrt{2}}(\hat{b}-\hat{b}^\dag), \nonumber\\
\end{eqnarray}
for the two modes $a$ and $b$, respectively. The EPR correlation is then defined as the total variance of the operators $\hat{x}_1-\hat{x}_2$ and $\hat{p}_1+\hat{p}_2$. The correlation for the EPR state \cite{epr}, which is a maximally entangled state, is $\Delta(\hat{x}_1-\hat{x}_2)^2$ = $\Delta(\hat{p}_1+\hat{p}_2)^2$ = 0. For a two-mode vacuum state, the EPR correlation is $\Delta(\hat{x}_1-\hat{x}_2)^2$ + $\Delta(\hat{p}_1+\hat{p}_2)^2$ = 2. For states with classical EPR correlation the variance difference is always greater than 2.

\begin{figure}
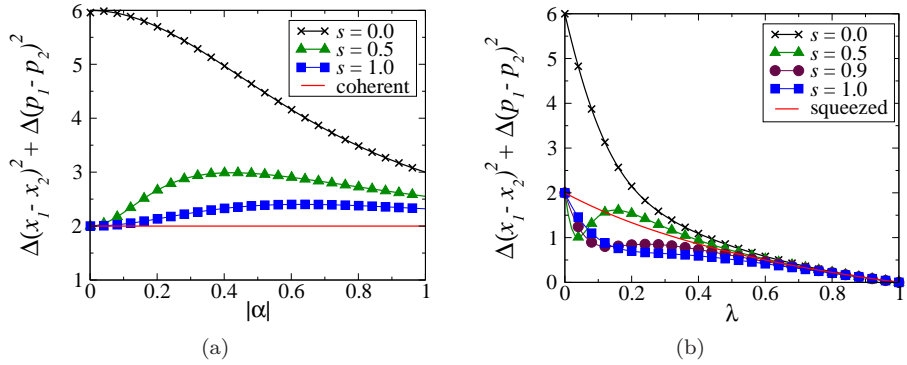

\begin{center}
\subfigure[]{\includegraphics[width=5.6cm]{Fig10a_EPR_2coh.eps}}\hspace{0.6cm}
\subfigure[]{\includegraphics[width=5.6cm]{Fig10b_EPR_2sqz.eps}}
\caption{Behaviour of EPR correlations, $\Delta(\hat{x}_1-\hat{x}_2)^2$ + $\Delta(\hat{p}_1+\hat{p}_2)^2$, for GSP operated (a) two-mode input coherent state, and (b) two-mode input squeezed state, with respect to varying (a) coherent field amplitude, and (b) $\lambda ~(=\tanh r)$ where $r$ is the squeezing parameter, for different values of the GSP operator parameter: $s$ = 0 (black crosses), $s$ = 0.5 (green triangles), $s$ = 0.9 (brown circles) [not shown in (a)], $s$ = 1.0 (blue squares), and unoperated states (red continuous line).}
\end{center}
\label{epr}
\end{figure}

Figure \ref{epr} gives us the variance of the EPR operator for the two-mode GSP operated coherent and squeezed state inputs. From the values of the variance, with changing field parameters, it is evident that the two-mode GSP operated coherent state is not EPR correlated at any field value. The variance for the two-mode non-operated coherent state is constant at 2.0. The EPR correlation patterns throw light into the teleportation fidelity plots of the two-mode coherent state [figure \ref{fid2cohthm}(a) and figure \ref{fidsqz1}(a)]. The maximum fidelity achieved for low $|\alpha|$, $F_{max}$ = 0.5, which is at the upper limit of teleportation using unentangled quantum channels ($F_{class}$=$\frac{1}{2}$). This is explained by the fact that the two-mode state is not EPR correlated for all $|\alpha|$ and has no quadrature entanglement. Hence, the GSP operation is unable to introduce phase-quadrature entanglement in the GSP operated coherent state even though low intensity states are two-mode entangled in the number state basis.

In comparison, figure \ref{epr}(b) shows that the two-mode squeezed state is strongly EPR correlated, with the GSP operation enhancing the correlation. This results in high teleportation fidelity above the cloning fidelity limit even for relatively low squeezing values ($\lambda$). The variances in the EPR quadrature operators are equal and behave as $\exp[-2r]$, where $r$ is the squeezing parameter. Hence, at maximum squeezing the GSP operated state is maximally EPR correlated.

Hence, we observe that non-Gaussian operations, such as the GSP operations, are able to introduce nonclassicality in optical states and thus create two-mode entanglement in the number state basis, in the low intensity regime, even when the initial states are coherent and thermal. However, these operations do not inject any entanglement in the phase quadrature or generate EPR correlations, which are crucial to gain advantage in CV quantum information protocols. Interestingly, for initial two-mode squeezed state inputs, which already contain significant quadrature correlations, the GSP operations are able to enhance EPR correlations to offer advantage in information protocols such as CV teleportation. Hence, the role of non-Gaussian operations in enhancing quantum resources is subject to the class of input states considered in the protocol. We further note that the two-mode entanglement generated in low intensity fields, under GSP operations, may play a significant role in discrete level information protocols.

\section{Discussion}
\label{discuss}

In this paper, we have used a generalised superposition of products of annihilation and creation operators to obtain nonclassical continuous variable field states. We have applied the GSP operator along with linear optical interaction on single- and two-mode CV states such as coherent, thermal and squeezed states to obtain highly entangled two-mode output states. The primary findings of the paper can be efficiently summarised in two logical parts. The first is the generation of highly entangled, two-mode discrete level  output states using the GSP operation. 
Upon application of the GSP protocol, the classical CV inputs, such as coherent and thermal states, can be converted to entangled two-mode states, in the low field-intensity limit. Manipulating the GSP parameter $s$ along with the field parameters, one can obtain highly entangled discrete photonic systems. Entangled photons are useful in a wide range of applications in quantum optics and are of fundamental importance in practical implementation of discrete level quantum information theory.

The second aspect of the study is the application of the GSP operated states as quantum channels in CV quantum teleportation. We observe that the GSP operations enhance average teleportation fidelity of teleporting single-mode coherent and squeezed states using two-mode entangled channels obtained from single- and two-mode squeezed state inputs. The average fidelity obtained is higher than the fidelity of cloning even at low squeezing regime and betters the performance of the non-operated two-mode squeezed channel. The optimised GSP operation outperforms most non-Gaussian operations with regard to fidelity enhancement in CV teleportation. Interestingly, the GSP operation does not enhance the classical fidelity when the two-mode GSP operated thermal and coherent channels are used. The maximum fidelity possible using the coherent and thermal channels is the classical fidelity that is achieved without any entangled resources. Hence, the nonclassical states introduced by GSP operations on coherent and thermal states are not resourceful for CV teleportation. This proves that not all two-mode entangled states generated under GSP operations are useful for CV information protocols. We show that this is due to the fact that GSP operations do not inject any nonclassicality in quadrature or EPR correlations, thus rendering these states useless for CV information protocols.
The results obtained may have interesting ramifications on noisy quantum channels. Such channels can be modelled by a linear characteristic function of a low intensity thermal state and an entangled two-mode squeezed channel. For the GSP operations considered in our study, and restrictive optimizations under the BK protocol, the average fidelity of teleportation with a squeezed channel is enhanced while that with a noisy thermal state is reduced. This could lead to interesting interplay between nonclassicality and noise in determining the overall teleportation fidelity.

The experimental advances in the generation of photon-added and photon-subtracted field states ensure that the advantages of non-Gaussian operations will be put to further applications in the future. In the light of such developments the importance of theoretical studies on non-Gaussian states is highly relevant.

\ack
HSD thanks University Grants Commission, Government of India for support under the Senior Research Fellowship scheme. AC thanks SERB, Department of Science and Technology, Government of India.

\begin{appendix}

\section{Characteristic function for arbitrary two-mode density matrix}
\label{app1}

The general expression of a symmetrically ordered characteristic function for a density matrix ($\rho$) in terms of its number state matrix elements, for a single-mode state, is given by \cite{Barnett}:
\begin{equation}
\chi(\eta)=\sum_{n,n'=0}^\infty \rho_{n,n'}\langle n|\hat{D}(\eta)|n'\rangle.
\label{a1}
\end{equation}
Generalizing the above relation for a two-mode density matrix, we obtain
\begin{equation}
\chi(\xi,\eta)=\sum_{n',n,m',m=0}^\infty \rho_{nm,n'm'}\langle n m|\hat{D}_a(\xi)\hat{D}_b(\eta)|n' m'\rangle .
\label{a2}
\end{equation}
To obtain the characteristic function, one needs to derive the inner product $\langle n m|\hat{D}_a(\xi)\hat{D}_b(\eta)|n' m'\rangle$. 
The general expression can be written as
\begin{eqnarray}
&& \langle n, m|\hat{D}_a(\xi)\hat{D}_b(\eta)|n-l, m-l'\rangle \nonumber\\
&=&\exp[-\frac{1}{2}(|\xi|^2+|\eta|^2)]\langle n, m|\exp(\xi\hat{a}^\dag)\exp(-\xi^*\hat{a})\exp(\eta\hat{b}^\dag)\exp(-\eta^*\hat{b})|n-l, m-l'\rangle \nonumber\\
&=& \exp[-\frac{1}{2}(|\xi|^2+|\eta|^2)]\langle n, m| \exp(\xi\hat{a}^\dag)\exp(-\xi^*\hat{a})\exp(\eta\hat{b}^\dag)\exp(-\eta^*\hat{b})\hat{a}^l\hat{b}^{l'}|n, m \rangle \nonumber\\
&& \times \sqrt{\frac{(n-l)!(m-l')!}{n!~m!}} \nonumber\\
&=&  \exp[-\frac{1}{2}(|\xi|^2+|\eta|^2)]\sqrt{\frac{(n-l)!(m-l')!}{n!~m!}}\left(-\frac{\partial}{\partial\xi^*}\right)^l\left(-\frac{\partial}{\partial\eta^*}\right)^{l'} \nonumber\\
&& \times\langle n, m|\exp(\xi\hat{a}^\dag)\exp(-\xi^*\hat{a})\exp(\eta\hat{b}^\dag)\exp(-\eta^*\hat{b})|n, m\rangle \nonumber\\
&=& ~ \exp[-\frac{1}{2}(|\xi|^2+|\eta|^2)]\sqrt{\frac{(n-l)!(m-l')!}{n!~m!}}(-\xi)^l (-\eta)^{l'}\left(-\frac{\partial}{\partial|\xi|^2}\right)^l\left(-\frac{\partial}{\partial|\eta|^2}\right)^{l'}\nonumber\\
&& \times 
\langle n, m|\exp(\xi\hat{a}^\dag)\exp(-\xi^*\hat{a})\exp(\eta\hat{b}^\dag)\exp(-\eta^*\hat{b})|n, m\rangle.\
\label{a3}
\end{eqnarray}
We now need to evaluate the term $\langle n, m|\exp(\xi\hat{a}^\dag)\exp(-\xi^*\hat{a})\exp(\eta\hat{b}^\dag)\exp(-\eta^*\hat{b})|n, m\rangle$ for all $m$, $n$.
\begin{eqnarray}
&& \langle n, m|\exp(\xi\hat{a}^\dag)\exp(-\xi^*\hat{a})\exp(\eta\hat{b}^\dag)\exp(-\eta^*\hat{b})|n, m\rangle \nonumber\\
&=& \langle n,m| \sum_{s=0}^\infty \frac{\xi^s\hat{a}^{\dag s}}{s!} \sum_{t=0}^\infty \frac{(-\xi^*)^{t}\hat{a}^{t}}{t!} \sum_{s'=0}^\infty \frac{\eta^{s'}\hat{b}^{\dag s'}}{s'!} \sum_{t'=0}^\infty \frac{(-\eta^*)^{t'}\hat{b}^{\dag t'}}{t'!}|n,m\rangle \nonumber\\
&=& \sum_{s,t,s',t'=0}^\infty \frac{\xi^s (-\xi^*)^{t}\eta^{s'}(-\eta^*)^{t'}}{s!~t!~s'!~t'!} \langle n,m|\hat{a}^{\dag s}\hat{b}^{\dag s'}\hat{a}^t\hat{b}^{t'}|n,m\rangle \nonumber\\
&=& \sum_{s,t,s',t'=0}^\infty \frac{\xi^s (-\xi^*)^{t}\eta^{s'}(-\eta^*)^{t'}}{s!~t!~s'!~t'!}
\sqrt{\frac{n!~m!}{(n-s)!(n-t)!}} \nonumber\\
&& \times \sqrt{\frac{n!~m!}{(m-s')!(m-t')!}}\underbrace{\langle n-s, m-s'|n-t, m-t'\rangle}{\delta_{st}\delta_{s't'}} \nonumber\\
&=& \sum_{s,s'=0}^\infty \frac{(|\xi|^2)^s(|\eta|^2)^{s'}}{(s!~s'!)^2}\left(\frac{n!}{(n-s)!}\right) \left(\frac{m!}{(m-s')!}\right) \nonumber\\
&=& L_n(|\xi|^2) L_m(|\eta|^2) ,
\label{a4}
\end{eqnarray}
where $L_n(x)$ are the Laguerre polynomials. 
Using relations (\ref{a3}) and (\ref{a4}), we get
\begin{eqnarray}
& & \langle n, m|\hat{D}_a(\xi)\hat{D}_b(\eta)|n-l, m-l'\rangle \nonumber\\ 
&=&  \exp[-\frac{1}{2}(|\xi|^2+|\eta|^2)]\sqrt{\frac{(n-l)!(m-l')!}{n!~m!}} \nonumber\\
&\times& (-\xi)^l (-\eta)^{l'}\left(-\frac{\partial}{\partial|\xi|^2}\right)^l\left(-\frac{\partial}{\partial|\eta|^2}\right)^{l'}L_n(|\xi|^2) L_m(|\eta|^2).
\label{a5}
\end{eqnarray}
%
Further,
\begin{equation}
\left(-\frac{\partial}{\partial x}\right)^l L_n(x) = L_n^l(x) ,
\label{a6}
\end{equation}
where $L_n^l(x)$ is the generalised or the associated Laguerre polynomials.
Hence, using relation (\ref{a6}) in (\ref{a2}), we can obtain the characteristic function of a two-mode state, provided its number state density matrix elements are known. This relation can be useful in cases where truncated number states are considered to represent the field.

\section{Characteristic function of GSP operated two-mode coherent state}
\label{app2}

The GSP operated two-mode coherent state interacting via a beam splitter can be written in the following way:
\begin{eqnarray}
|\phi\rangle_{ab} &=& \hat{\mathcal{B}}_{ab}\{s_1s_2+s_1(s_2+t_2)\hat{b}^\dag\hat{b}+s_2(s_1+t_1)\hat{a}^\dag\hat{a}+ (s_1+t_1)\nonumber\\
 & & \times(s_2+t_2)\hat{a}^\dag\hat{b}^\dag\hat{a}\hat{b}\}|\alpha,\beta \rangle .
\label{b1}
\end{eqnarray}
The characteristic function can be written as
\begin{eqnarray}
\chi(\xi,\eta)=~_{ab}\langle \phi|D_a(\xi)D_b(\eta)|\phi \rangle_{ab} .
\label{b3}
\end{eqnarray}
An important step in the calculation is the evaluation of the following term:
\begin{eqnarray}
& & \langle \alpha, \beta|\hat{\mathcal{B}}_{ab}^\dag D_a(\xi)D_b(\eta)\hat{\mathcal{B}}_{ab}|\alpha,\beta\rangle \nonumber\\
 & = & \langle \alpha, \beta| D_a(\xi/2)D_b(\xi/2)D_a(-\eta/2)D_b(\eta/2)|\alpha,\beta\rangle ,
\label{b4}
\end{eqnarray}
where we have used the properties of the beam-splitter and the displacement operator. The relation can be further simplified using the following identities:
\begin{eqnarray}
D_a\left(\frac{\xi}{2}\right)D_a\left(-\frac{\eta}{2}\right) &=& D_a\left(\frac{\xi-\eta}{2}\right)\exp[\frac{1}{4}(-\xi\eta^*+\xi^*\eta)] , \nonumber\\
D_b\left(\frac{\xi}{2}\right)D_b\left(\frac{\eta}{2}\right) &=& D_b\left(\frac{\xi+\eta}{2}\right)\exp[\frac{1}{4}(\xi\eta^*-\xi^*\eta)] .
\end{eqnarray}
Using (\ref{b4}) in relation (\ref{b3}), we get
\begin{eqnarray}
& & \langle \alpha, \beta|\hat{\mathcal{B}}_{ab}^\dag D_a(\xi)D_b(\eta)\hat{\mathcal{B}}_{ab}|\alpha,\beta\rangle\nonumber\\
 &=& \langle \alpha, \beta|D_a\left(\frac{\xi-\eta}{2}\right)D_b\left(\frac{\xi+\eta}{2}\right)|\alpha,\beta\rangle \nonumber\\
&=& \exp[-\frac{1}{4}((\xi-\eta)^*\alpha-(\xi-\eta)\alpha^*)]\exp[-\frac{1}{4}((\xi+\eta)^*\beta-(\xi+\eta)\beta^*)] \nonumber\\
&& \times \langle \alpha,\beta|\frac{\xi-\eta}{2}+\alpha,\frac{\xi+\eta}{2}+\beta\rangle \nonumber\\
&=& \exp[-\frac{1}{2}(X^*\alpha-X\alpha^*)]\exp[-\frac{1}{2}(Y^*\beta-Y\beta^*)]\langle \alpha,\beta|X+\alpha,Y+\beta\rangle \nonumber\\
&=& \exp[-\frac{1}{2}(|X|^2+|Y|^2)]\exp(\alpha^*X-\alpha X^*)\exp(\beta^*Y-\beta Y^*) ,
\label{b5}
\end{eqnarray}
where $X=\frac{\xi-\eta}{\sqrt{2}}$, and $Y=\frac{\xi+\eta}{\sqrt{2}}$. Let $\langle \alpha, \beta|\hat{\mathcal{B}}_{ab}^\dag D_a(\xi)D_b(\eta)\hat{\mathcal{B}}_{ab}|\alpha,\beta\rangle$= $\mathcal{H}$. Proceeding as in (\ref{b5}), we can derive the following relations:

\begin{eqnarray}
& & \langle \alpha, \beta|\hat{\mathcal{B}}_{ab}^\dag D_a(\xi)D_b(\eta)\hat{\mathcal{B}}_{ab}\hat{a}^\dag\hat{a}|\alpha,\beta\rangle\nonumber\\ &=& \langle \alpha, \beta|\hat{\mathcal{B}}_{ab}^\dag D_a(\xi)D_b(\eta)\hat{\mathcal{B}}_{ab}|\alpha,\beta\rangle (|\alpha|^2-\alpha X^*) \equiv \mathcal{H}c_1 \nonumber\\\nonumber\\
& & \langle \alpha, \beta|\hat{a}^\dag\hat{a}\hat{\mathcal{B}}_{ab}^\dag D_a(\xi)D_b(\eta)\hat{\mathcal{B}}_{ab}|\alpha,\beta\rangle\nonumber\\ &=& \langle \alpha, \beta|\hat{\mathcal{B}}_{ab}^\dag D_a(\xi)D_b(\eta)\hat{\mathcal{B}}_{ab}|\alpha,\beta\rangle (|\alpha|^2+\alpha^* X) \equiv \mathcal{H}c_2 \nonumber\\\nonumber\\
& & \langle \alpha, \beta|\hat{a}^\dag\hat{a}\hat{\mathcal{B}}_{ab}^\dag D_a(\xi)D_b(\eta)\hat{\mathcal{B}}_{ab}\hat{a}^\dag\hat{a}|\alpha,\beta\rangle\nonumber\\ &=& \langle \alpha, \beta|\hat{\mathcal{B}}_{ab}^\dag D_a(\xi)D_b(\eta)\hat{\mathcal{B}}_{ab}|\alpha,\beta\rangle |\alpha|^2(1+(\alpha+X)(\alpha^*- X^*))\nonumber\\
&\equiv& \mathcal{H}c_3 .
\label{b6}
\end{eqnarray}
Let $d_i$ ($i$ = 1, 2, 3) be the inner products corresponding to $\hat{b}^{\dagger} \hat{b}$ (similar to $c_i$ in (\ref{b6})):
\begin{eqnarray}
d_1 &=& |\beta|^2-\beta Y^* \nonumber\\
d_2 &=& |\beta|^2+\beta^* Y \nonumber\\
d_3 &=& |\beta|^2(1+(\beta+Y)(\beta^*-Y^*)) .
\label{b7}
\end{eqnarray}
Using relations (\ref{b5}), (\ref{b6}), and (\ref{b7}), we can write the normalised characteristic function for a GSP operated two-mode coherent state interacting via a symmetric beam-splitter as
\begin{eqnarray}
& & \chi(\xi,\eta)\nonumber\\ &=& (s_1^2s_2^2 + s_1 s_2^2(s_1+t_1)(c_1+c_2)+s_2 s_1^2(s_2+t_2)(d_1+d_2) \nonumber\\
&& +(s_1+t_1)^2(s_2+t_2)^2 c_3d_3+s_1s_2(s_1+t_1)(s_2+t_2) \nonumber\\
&& \times(c_1d_1+c_2d_2+c_1d_2+c_2d_1)+s_2(s_1+t_1)^2c_3+ s_1(s_2+t_2)^2d_3 \nonumber\\
&& +s_1(s_1+t_1)(s_2+t_2)^2 d_3(c_1+c_2)+ s_2(s_1+t_1)^2(s_2+t_2)c_3\nonumber\\
& & \times(d_1+d_2))\mathcal{H}\times\frac{1}{\mathcal{N}} ,
\end{eqnarray}
where $\mathcal{N}$ is the normalization constant given by
\\\\
\begin{eqnarray}
& &\mathcal{N}\nonumber\\ &=& s_1^2s_2^2+s_1^2(s_2+t_2)^2(|\beta|^2+|\beta|^4)+s_2^2(s_1+t_1)^2(|\alpha|^2+|\alpha|^4) \nonumber\\
&& +(s_1+t_1)^2(s_2+t_2)^2(|\beta|^2+|\beta|^4)(|\alpha|^2+|\alpha|^4)+2s_1^2s_2(s_2+t_2)|\beta|^2 \nonumber\\
&& +2s_2^2s_1(s_1+t_1)|\alpha|^2+4s_1s_2(s_1+t_1)(s_2+t_2)|\alpha|^2|\beta|^2 + 2s_1(s_1+t_1) \nonumber\\
&& \times (s_2+t_2)^2|\alpha|^2(|\beta|^2+|\beta|^4)+2s_2(s_2+t_2)(s_1+t_1)^2|\beta|^2 \nonumber\\
&& \times (|\alpha|^2+|\alpha|^4) .
\end{eqnarray}

\section{Characteristic function of GSP operated two-mode squeezed state}
\label{app3}

The GSP operated two-mode squeezed state can be written in the following way:
\begin{eqnarray}
|\psi\rangle_{ab} &=& \{s_1t_1+s_1(s_2+t_2)\hat{b}^\dag\hat{b}+s_2(s_1+t_1)\hat{a}^\dag\hat{a} + (s_1+t_1)\nonumber\\
& & \times(s_2+t_2)\hat{a}^\dag\hat{b}^\dag\hat{a}\hat{b}\}\hat{S}(r)|0\rangle_a|0\rangle_b \nonumber\\
&=& \hat{U}_{sup}~\hat{S}(r)|0\rangle_a|0\rangle_b ,
\label{c1}
\end{eqnarray}
where $\hat{S}(r)=\exp[r~(\hat{a}^{\dag}\hat{b}^\dag-\hat{a}\hat{b})]$ is the two-mode squeezing operator, and $r$ is the squeezing parameter. The characteristic function can be written as
\begin{eqnarray}
\chi(\xi,\eta) =_{ab}\langle 00|\hat{S}^\dag(r) \hat{U}_{sup}^\dag D_a(\xi)D_b(\eta) \hat{U}_{sup}\hat{S}(r)|00\rangle_{ab} .
\label{c2}
\end{eqnarray}
We first calculate the following term:
\begin{eqnarray}
& & \langle 00|\hat{S}^\dag(r) D_a(\xi)D_b(\eta) \hat{S}(r)|00\rangle\nonumber\\
 &=& \langle 00|\hat{S}^\dag(r) \exp[\xi\hat{a}^\dag-\xi^*\hat{a}]\exp[\eta\hat{b}^\dag-\eta^*\hat{b}] \hat{S}(r)|00\rangle ,
\label{c3}
\end{eqnarray}
where we have dropped the mode indices $a$, $b$ from the states, and used the definition of the displacement operator. Using the fact that
\begin{eqnarray}
\hat{S}^\dag(r)\hat{a}^\dag \hat{S}(r) &=& \cosh r~\hat{a}^\dag + \sinh r~ \hat{b} , \nonumber\\
\hat{S}^\dag(r)\hat{b}^\dag\hat{S}(r) &=& \cosh r~\hat{b}^\dag + \sinh r~ \hat{a} ,
\label{c4}
\end{eqnarray}
and expanding relation (\ref{c3}), we get
\begin{eqnarray}
& & \langle 00|\hat{S}^\dag(r) D_a(\xi)D_b(\eta) \hat{S}(r)|00\rangle\nonumber\\
&=&\langle 00|\hat{S}(r)^\dag \exp[\xi\hat{a}^\dag-\xi^*\hat{a}]\exp[\eta\hat{b}^\dag-\eta^*\hat{b}] \hat{S}(r)|00\rangle \nonumber\\
&=& \langle 00| \exp [\xi(\cosh r~\hat{a}^\dag+\sinh r~\hat{b})-\xi^*(\cosh r~\hat{a}+\sinh r~\hat{b}^\dag)] \nonumber\\
&& \times \exp[\eta(\cosh r~\hat{b}^\dag+\sinh r~\hat{a})-\eta^*(\cosh r~\hat{b}+\sinh r~\hat{a}^\dag)]|00\rangle \nonumber\\
&=& \langle 00| D_a(\xi')D_b(\eta')|00\rangle\nonumber\\
& = &\exp[-\frac{1}{2}(|\xi'|^2+|\eta'|^2)] ,
\label{c5}
\end{eqnarray}
where $\xi' = \xi \cosh r-\eta^* \sinh r$, and $\eta'= \eta \cosh r-\xi^*\sinh r$. Let $\langle 00|\hat{S}^\dag(r) D_a(\xi)D_b(\eta) \hat{S}(r)|00\rangle \equiv \mathcal{L}$. The characteristic function of the GSP operated two-mode squeezed state can be calculated using the results (\ref{c4}) and (\ref{c5}). Let us consider some general terms to be calculated to obtain the characteristic function.
\begin{eqnarray}
& & \langle 00| \hat{S}^\dag(r)D_a(\xi)D_b(\eta)\hat{a}^\dag\hat{a}~\hat{S}(r)|00\rangle\nonumber\\
 &=& (\sinh^2 r+ (\xi' \eta')^*\cosh r\sinh r)\mathcal{L} = (e_0+e_1(\xi' \eta')^*)\mathcal{L}\nonumber\\
 &\equiv& g_0\mathcal{L} \nonumber\\ \nonumber\\
& & \langle 00| \hat{S}^\dag(r)D_a(\xi)D_b(\eta)\hat{a}^\dag\hat{a}\hat{a}^\dag\hat{a}~\hat{S}(r)|00\rangle\nonumber\\
&=& ((\xi'^*\eta'^*)^{2}\sinh^2 r\cosh^2r+(\xi'\eta')^*(\sinh r\cosh^3 r+3\sinh^3 r\cosh r)\nonumber\\
& & + (\sinh^2 r+\sinh^4 r))\mathcal{L} \nonumber\\
&\equiv& ((\xi'^*\eta'^*)^{2}e_1^2+(\xi'\eta')^*(3~e_0e_1+\frac{e_1^3}{e_0}) + (e_0^2+e_1^2))\mathcal{L} \nonumber\\
&\equiv& ((\xi'^*\eta'^*)^{2}(f_2/2)+(\xi'\eta')^*f_1+f_0)\mathcal{L}  \nonumber\\
&\equiv& g_1\mathcal{L} \nonumber\\ \nonumber\\
& & \langle 00| \hat{S}^\dag(r)\hat{a}^\dag\hat{a}D_a(\xi)D_b(\eta)\hat{a}^\dag\hat{a}~\hat{S}(r)|00\rangle\nonumber\\
&=& e_0(e_0+e_1(\xi'\eta')^*)+ e_1(e_0(\xi'\eta') +e_1(1-|\xi'|^2-|\eta'|^2+|\xi'|^2|\eta'|^2) \mathcal{L} \nonumber\\
&\equiv& g_2\mathcal{L} \nonumber\\\nonumber\\
& & \langle 00| \hat{S}^\dag(r)\hat{a}^\dag\hat{a}D_a(\xi)D_b(\eta)\hat{a}^\dag\hat{a}\hat{a}^\dag\hat{a}~\hat{S}(r)|00 \rangle\nonumber\\
&=& f_0(e_0+e_1(\xi'\eta'))+f_1(e_0(\xi'\eta')^* +e_1(1-|\xi'|^2-|\eta'|^2+|\xi'|^2|\eta'|^2))\nonumber\\
&&+ f_2\left(\frac{e_0 (\xi'^*\eta'^*)^2}{2}+ e_1(\xi'\eta')^*\left(2-|\xi'|^2-|\eta'|^2+\frac{|\xi'|^2|\eta'|^2}{2}\right)\right)\mathcal{L} \nonumber\\
&\equiv& g_3 \mathcal{L}\nonumber\\\nonumber\\
& & \langle 00| \hat{S}^\dag(r)\hat{a}^\dag\hat{a}\hat{a}^\dag\hat{a}D_a(\xi)D_b(\eta)\hat{a}^\dag\hat{a}\hat{a}^\dag\hat{a}~\hat{S}(r)|00\rangle\nonumber\\
 &=& f_0\left(f_0+f_1(\xi'\eta')^*+f_2\frac{(\xi'^*\eta'^*)^2}{2}\right)+ f_1 (f_0(\xi'\eta') + f_1(1-|\xi|^2-|\eta|^2 \nonumber\\
&& + |\xi|^2|\eta|^2)+f_2 (\xi'\eta')^*\left(2-|\xi'|^2-|\eta'|^2 +\frac{|\xi'|^2|\eta'|^2}{2}\right))+ f_2\left(f_0\frac{(\xi'\eta')^2}{2}\right.\nonumber\\
&&\left.+ f_1(\xi'\eta')\left(2- |\xi'|^2 -|\eta'|^2 + \frac{|\xi'|^2|\eta'|^2}{2}\right) + f_2\left(2 - 4|\xi'|^2 + |\xi'|^4 \right.\right.  \nonumber\\
&& \left.\left.- 2|\xi'|^2|\eta'|^4 + 8|\xi'|^2|\eta'|^2 +\frac{|\xi'|^4|\eta'|^4}{2} - 2|\eta'|^2|\xi'|^4+ |\eta'|^4 - 4|\eta'|^2 \right)\right) \mathcal{L}\nonumber\\
& &   \equiv g_4\mathcal{L}.\nonumber\\
\label{c6}
\end{eqnarray}
In terms of the relations (\ref{c6}), the characteristic function can be written as
\begin{eqnarray}
&&\chi(\xi,\eta)\nonumber\\ &=& (s_1^2s_2^2+\{s_1^2s_2(s_2+t_2)+s_1s_2^2(s_1+t_1)\}(g_0+g^*_0) \nonumber\\
&& + s_1s_2(s_1+t_1)(s_2+t_2)(g_1+g_1^*+2g_2)+\{s_1(s_1+t_1) \nonumber\\
&& \times (s_2+t_2)^2+s_2(s_2+t_2)(s_1+t_1)^2\}(g_3+g_3^*)+\{s_1^2(s_2+t_2)^2 \nonumber\\
&& + s_2^2(s_1+t_1)^2 +(s_1+t_1)^2(s_2+t_2)^2 g_4)~ \mathcal{L}\times \frac{1}{\mathcal{N}} ,
\label{c7}
\end{eqnarray}
where $\mathcal{N}$ is the normalization constant given by
\begin{eqnarray}
&&\mathcal{N}\nonumber\\ &=& s_1^2s_2^2+2e_0\{s_1^2s_2(s_2+t_2)+s_1s_2^2(s_2+t_2)\} \nonumber\\
&& + \{4s_1s_2(s_1+t_1)(s_2+t_2)+s_1^2(s_2+t_2)^2+s_2^2(s_1+t_1)^2\} \nonumber\\
&& \times (e_0^2+e_1^2)+2\{s_1(s_1+t_1)(s_2+t_2)^2 +s_2(s_1+t_1)^2(s_2+t_2)\} \nonumber\\
&& \times (f_0e_0+f_1e_1)+(s_1+t_1)^2(s_2+t_2)^2(f_0^2+f_1^2+f_2^2).
\label{c8}
\end{eqnarray}
The terms $e_i, f_i$ and $g_i$ have been defined in relation (\ref{c6}).
\end{appendix}

\end{document}